\documentclass[preprint,sort&compress,12pt]{elsarticle}
\usepackage{amssymb}
\usepackage{amsthm}
\usepackage{amsmath}
\usepackage{mathtools}
\usepackage{mathrsfs}
\usepackage{algorithm}
\usepackage[algo2e]{algorithm2e}
\usepackage{algpseudocode}
\usepackage{array}
\usepackage{multirow}
\usepackage{listings}
\usepackage{tabu}
\usepackage{booktabs}
\usepackage{enumerate}
\usepackage{lineno}
\usepackage{fullpage}
\usepackage{xcolor}
\usepackage[colorinlistoftodos]{todonotes}
\usepackage{bm}
\usepackage{bbm}
\usepackage[colorlinks=true]{hyperref}
\usepackage{url}
\usepackage{textcomp}
\usepackage{gensymb}
\usepackage{soul}
\usepackage{graphicx}
\usepackage{subfig}
\usepackage{float}
\usepackage{caption}
\usepackage{tikz}
\usetikzlibrary{shapes}

\theoremstyle{definition}

\theoremstyle{remark}

\biboptions{numbers,comma,round,square}
\graphicspath{ {./figs/} }
\journal{Elsevier}

\newcommand\blfootnote[1]{%
	\begingroup
	\renewcommand\thefootnote{}\footnote{#1}%
	\addtocounter{footnote}{-1}%
	\endgroup
}

\begin{document}
\makeatletter
\def\ps@pprintTitle{%
  \let\@oddhead\@empty
  \let\@evenhead\@empty
  \let\@oddfoot\@empty
  \let\@evenfoot\@oddfoot
}
\makeatother
\begin{frontmatter}
%\tableofcontents
% \listoftodos

%% Title, authors and addresses
 \title{\textbf{CoNFiLD}: \textbf{Co}nditional \textbf{N}eural \textbf{Fi}eld \textbf{L}atent \textbf{D}iffusion Model Generating Spatiotemporal Turbulence}

%% use the tnoteref command within \title for footnotes;
%% use the tnotetext command for the associated footnote;
%% use the fnref command within \author or \address for footnotes;
%% use the fntext command for the associated footnote;
%% use the corref command within \author for corresponding author footnotes;
%% use the cortext command for the associated footnote;
%% use the ead command for the email address,
%% and the form \ead[url] for the home page:
%%
%% \title{Title\tnoteref{label1}}
%% \tnotetext[label1]{}
%% \author{Name\corref{cor1}\fnref{label2}}
%% \ead{email address}
%% \ead[url]{home page}
%% \fntext[label2]{}
%% \cortext[cor1]{}
%% \address{Address\fnref{label3}}
%% \fntext[label3]{}

%% use optional labels to link authors explicitly to addresses:
%% \author[label1,label2]{<author name>}
%% \address[label1]{<address>}
%% \address[label2]{<address>}

\author[ndAME]{Pan Du\corref{contrib}}
\author[ndAME]{Meet Hemant Parikh\corref{contrib}}
\author[ndAME]{Xiantao Fan}
\author[ndAME]{Xin-Yang Liu}
\author[ndAME,Energy,Lucy]{Jian-Xun Wang\corref{corr}}

\address[ndAME]{Department of Aerospace and Mechanical Engineering, University of Notre Dame, Notre Dame, IN, USA}
\address[Lucy]{Lucy Family Institute for Data \& Society, University of Notre Dame, Notre Dame, IN, USA}
\address[Energy]{Center for Sustainable Energy (ND Energy), University of Notre Dame, Notre Dame, IN, USA}

\cortext[contrib]{PD and MHP contributed equally}

\cortext[corr]{Corresponding author. Tel: +1 540 3156512}
\ead{jwang33@nd.edu}

\begin{abstract}
This study introduces the Conditional Neural Field Latent Diffusion (CoNFiLD) model, a novel generative learning framework designed for rapid simulation of intricate spatiotemporal dynamics in chaotic and turbulent systems within three-dimensional irregular domains. Traditional eddy-resolved numerical simulations, despite offering detailed flow predictions, encounter significant limitations due to their extensive computational demands, restricting their applications in broader engineering contexts. In contrast, deep learning-based surrogate models promise efficient, data-driven solutions. However, their effectiveness is often compromised by a reliance on deterministic frameworks, which fall short in accurately capturing the chaotic and stochastic nature of turbulence. The CoNFiLD model addresses these challenges by synergistically integrating conditional neural field encoding with latent diffusion processes, enabling the memory-efficient and robust probabilistic generation of spatiotemporal turbulence under varied conditions. Leveraging Bayesian conditional sampling, the model can seamlessly adapt to a diverse range of turbulence generation scenarios without the necessity for retraining, covering applications from zero-shot full-field flow reconstruction using sparse sensor measurements to super-resolution generation and spatiotemporal flow data restoration. Comprehensive numerical experiments across a variety of inhomogeneous, anisotropic turbulent flows with irregular geometries have been conducted to evaluate the model's versatility and efficacy, showcasing its transformative potential in the domain of turbulence generation and the broader modeling of spatiotemporal dynamics. \blfootnote{Videos of all numerical experiments can be found at \href{https://sites.nd.edu/jianxun-wang/confild-conditional-neural-field-latent-diffusion-model-generating-spatiotemporal-turbulence/}{https://sites.nd.edu/jianxun-wang/animations}}

\end{abstract}

\begin{keyword}
%% keywords here, in the form: keyword \sep keyword
  Generative AI \sep Surrogate Modeling \sep Chaotic Systems \sep Inverse Problems \sep Bayesian Learning \sep Data Assimilation
\end{keyword}
\end{frontmatter}

\nolinenumbers

\section{Introduction}
\label{sec:intro}

Turbulent flows, characterized by their inherent chaotic and multiscale nature, are a central subject in the study of fluid dynamics, essential for understanding phenomena in diverse areas such as aerospace, oceanography, and combustion. Traditionally, simulating these complex spatiotemporal behaviors has relied on first-principle eddy-resolving methods like Direct Numerical Simulation (DNS) and Large Eddy Simulation (LES), which require numerically solving the governing partial differential equations (PDEs) for fluid flows. While these methods offer detailed insights, their application is largely limited by significant computational demands. The fine-scale spatiotemporal resolution required by DNS and LES to accurately capture the wide range of space and time scales in turbulence structure results in substantial computational loads, making them impractical for most engineering applications.

The rapid advancements in machine/deep learning (ML/DL) have profoundly influenced computational fluid dynamics (CFD)~\cite{brunton2020machine,vinuesa2022enhancing}, bringing a fresh and innovative dimension to the field, marked by recent developments such as advanced DL-based discretization~\cite{bar2019learning,kochkov2021machine}, data-driven closure modeling~\cite{wang2017physics,maulik2019subgrid,shankar2023differentiable}, accelerated CFD solving processes~\cite{tompson2017accelerating}, and the differentiable hybrid neural modeling, a framework that unifies conventional CFD and DL through differentiable programming~\cite{liu2024multi,fan2024differentiable}. Moreover, DL has become instrumental in developing rapid surrogate or reduced-order models, offering efficient alternatives to computationally-intensive numerical solvers for emulating complex spatiotemporal dynamics. These models, often built on autoregressive learning architectures, are adept at predicting future flow states based on previous conditions, relying on temporal correlations learned from training data. An important aspect of these models is the integration of dimensionality reduction techniques, such as Proper Orthogonal Decomposition (POD) and Convolutional Neural Network (CNN) autoencoders, with sequence neural networks, e.g., Long-short Term Memory (LSTM) and transformers. Notable examples include the convolutional autoencoder-based autoregressive learning models by Fukami and co-workers for inflow turbulence synthesis and super-resolution~\cite{fukami2019synthetic,fukami2019super,fukami2021machine}, and the work of Yousif et al.~\cite{yousif2022physics}, who combined CNN autoencoders with LSTM networks, further advancing these models with adversarial training and attention mechanisms~\cite{yousif2023transformer}. To effectively handle unstructured flow data within irregular domains, Graph Neural Networks (GNN)-based autoencoder coupled with temporal attention models has been proposed and shown the effectiveness~\cite{han2022predicting}. Despite the promise, challenges remain, particularly in the turbulence regime. The deterministic nature of these ML-based surrogate models often inadequately captures the stochastic behavior inherent in turbulent flows. These models, largely relying on autoregressive architectures, are able to learn the complex distribution of turbulence, limiting their capacity to produce stochastic flow realizations. This can result in substantial deviations in long-term predictions, as the chaotic nature of turbulence magnifies the impact of even minor inaccuracies or perturbations. Furthermore, there is a risk of cumulative error propagation in these models, potentially undermining the robustness and reliability of their long-term forecasting capabilities. 

Generative AI, rooted in probabilistic learning and statistical inference, offers a promising direction to overcome these limitations. These models are capable of learning the complex probabilistic distributions within datasets, allowing for the generation of new data samples that statistically resemble the training sets. In the context of turbulence simulation, generative models are particularly valuable as they can capture the multi-scale and stochastic characteristics of turbulence, thereby enabling the synthesis of instantaneous flow field realizations that align with the statistical characteristics observed in real-world turbulent data. The recent surge in deep generative models for turbulence, primarily driven by Generative Adversarial Networks (GANs), underscores their potential and promise. GANs operate through a dynamic interplay between a generator, which creates synthetic turbulent data, and a discriminator, which distinguishes between synthetic and real data. This iterative adversarial process refines the generator's output, aiming for convergence to the actual data distribution. Variants like Wasserstein GAN (WGAN), conditional GAN (cGAN), deep convolutional GAN (DCGAN), super-resolution GAN (SRGAN), and cycle-consistent GAN (CycGAN) have been adapted for specific tasks, such as turbulence generation~\cite{stengel2020adversarial,drygala2022generative}, super-resolution~\cite{deng2019super,kim2021unsupervised,guemes2021coarse,yu2022three}, and data inpainting~\cite{buzzicotti2021reconstruction}. However, the primary limitation of these models is their focus on single-snapshot generation, as they are trained on isolated flow snapshots without temporal coherence, which restricts their ability to synthesize spatiotemporal turbulence. Attempts to integrate GANs with sequential networks have been made, but these often result in GANs acting as deterministic encoders, not fully exploiting their stochastic generation capabilities~\cite{yousif2023transformer}. Only a few studies, such as TempoGAN by Xie et al.~\cite{xie2018tempogan} and WGAN-RNN model by Kim and Lee~\cite{kim2020deep}, have leveraged GANs for stochastic turbulence generation. While GANs have shown potential for turbulence synthesis, they often face significant challenges:  their training is notoriously challenging due to the oscillatory behavior between the generator and discriminator components~\cite{gui2021review}. Additionally, they are susceptible to ``mode collapse'', a limitation that results in a reduced diversity of output in the generated simulations~\cite{bau2019seeing}. These factors critically impede their efficacy in accurately modeling complex turbulent dynamics. In addition to GANs, normalizing flows (NFlows) have also been explored for turbulence generation. These models stand out for their ability to directly model complex data distributions through a series of invertible, differentiable transformations, which is particularly useful in emulating intricate dynamics like turbulence. Geneva and Zabaras~\cite{geneva2020multi} have utilized NFlows for super-resolving Very Large Eddy Simulation (VLES) data, and Sun et al.~\cite{sun2023unifying} developed a sequential NFlows model integrating GNN-autoencoding and attention mechanism to synthesize instantaneous backward-facing-step flows. However, NFlows have the known scalability issue due the complexity of computing Jacobians in transformations, making them infeasible to handle real-world 3D turbulence data. 

Diffusion models have recently advanced the field of generative modeling, outperforming GANs and NFlows in a variety of computer vision tasks~\cite{dhariwal2021diffusion,croitoru2023diffusion,yang2023diffusion}. These models are uniquely characterized by their progressive approach of transforming data from a simple distribution into a complex one. This is achieved by initially introducing noise into the dataset and then systematically denoising it through deep neural networks (DNNs). There are two primary categories of diffusion models: denoising diffusion probabilistic models (DDPMs)~\cite{ho2020denoising} and score-based diffusion models~\cite{song2019generative}, both of which can be unified within the stochastic differential Equation (SDE)-based framework~\cite{song2020score}. The advantages of diffusion models are manifold, including the ease of training, the capability of capturing multi-scale features, and their proficiency in conditional generation, particularly within a Bayesian framework. While diffusion models have recently shown significant success in fields like image generation and super-resolution~\cite{dhariwal2021diffusion,gao2023implicit}, their application in turbulence simulation represents an emerging and largely uncharted domain. Recent studies have explored the use of DDPMs in super-resolution or inpainting of turbulence data~\cite{shu2023physics,wan2023debias,li2023multi}. However, these initial works mainly focused on single-snapshot generations, typically in 2D Komogrov flows, homogenous and isotropic in nature. Most recently, Gao et al.~\cite{gao2023bayesian} have taken a leap forward with the development of a Bayesian conditional diffusion model for spatiotemporal turbulence generation. This model has showcased its capability to stochastically generate the temporal evolution of complex, wall-bounded turbulence in a variety of conditions, including URANS super-fidelity, auto-regressive generation, and super-resolution generation. However, the foundational architecture of this model, VideoDiffusion~\cite{ho2022video}, utilizes 3D convolution in physical spatiotemporal space, encountering scalability and efficiency challenges. This limitation confines its application to small-scale 2D spatial fields with a limited temporal extent of the generated segments. Furthermore, the model's backbone architecture, a CNN-based 3D U-Net, inherently requires regular domains with uniform grids, posing a limitation in handling complex, irregular geometries with unstructured grids, which are prevalent in CFD, thereby restricting its adaptability to a broader range of real-world turbulence simulation scenarios.   
    
In this work, we proposed a conditional neural field latent diffusion (CoNFiLD) model, innovatively designed for efficiently generating complex spatiotemporal dynamics of chaotic and turbulent systems across diverse conditions, addressing both regular and irregular geometrical configurations. Distinct from the majority of existing literature that focuses on single-snapshot (image) generation, CoNFiLD emphasizes capturing the probabilistic distribution of time-evolving turbulent flow sequences, operating as a sophisticated stochastic spatiotemporal process, allowing for an effective generation of new instantaneous flow realizations through random sampling under a variety of conditions. The proposed model synergistically integrates conditional neural field (CNF) techniques with a latent probabilistic diffusion model, enabling forward and reverse diffusion process being operated in the CNF-encoded latent space. This innovative architecture leverages the advantage and effectiveness of CNF for meshless nonlinear dimension reduction, which have been demonstrated in recent literature~\cite{pan2023neural,yin2022continuous}, ensuring robust performance in diverse geometrical configurations and scalable applications. By significantly improving the scalability and efficiency of both offline training and online generation, this work overcomes the limitations of the previous model: scalability constraints and uniform grid requirements~\cite{gao2023bayesian}. Moreover, the proposed CoNFiLD is also featured for its zero-shot conditional generation by leveraging the Bayesian formulation and differentiable programming, thereby eliminating the need for retraining when adapting to new flow conditions. The generative learning capability of our CoNFiLD model has been showcased through its application to a wide range of real-world 3D turbulent flow cases, including scenarios with wall-bounded turbulence, flow separations, and intricate 3D geometries. Remarkably versatile, the unconditionally trained CoNFiLD model can be directly applied for conditional generation tasks without the need of re-training. These applications span from reconstructing full-field spatiotemporal flows from sparse sensor data to generating super-resolution spatiotemporal flows and restoring corrupted flow data. This work represents a significant contribution to the field of spatiotemporal generative modeling and turbulence simulation, offering a comprehensive and efficient solution for generating realistic, complex instantaneous turbulent flows in various scenarios. To the best of the authors' knowledge, this study represents the first development of a neural field encoded latent diffusion model for the 4D generation of spatiotemporal dynamics in chaotic and turbulent systems with complex, irregular domains.

The remainder of this paper is structured as follows: Section~\ref{sec:meth} elaborates on the proposed CoNFiLD framework and its mathematical details. Section~\ref{sec:result} provides a comprehensive set of numerical experiments to evaluate and demonstrate CoNFiLD's generative capabilities across various wall-bounded turbulence scenarios. The computational efficiency, memory usage, and scalability comparisons of CoNFiLD with other methodologies are discussed in Section~\ref{sec:discussion}. The paper is summarized in Section~\ref{sec:conclusion}, outlining the main findings and contributions.

\section{Methodology}
\label{sec:meth}
%We describe the workflow the CoNFiLD model for spatiotemporal flow field generation in this section. We introduce overall architecture of CoNFiLD in Section ~\ref{sec:overview} with illustration (Figure ~\ref{fig:method}) and specify the major components- the encoding module and the latent diffusion module in Section ~\ref{sec:encoding} and ~\ref{sec:diffusion}, respectively. Finally in Section ~\ref{sec:uncon generation}, we describe the unconditional generation process and propose several innovative and effective ways of utilizing our model across a broad range of practical situations, demonstrating its versatility in addressing diverse challenges and opportunities.

\subsection{Overview of CoNFiLD generative learning framework}
\label{sec:overview}

\begin{figure}[t!]
    % \centering
    % \includegraphics[width=1.0\textwidth]{./figure/methodv4-small.pdf}
    \includegraphics[width=1.0\textwidth]{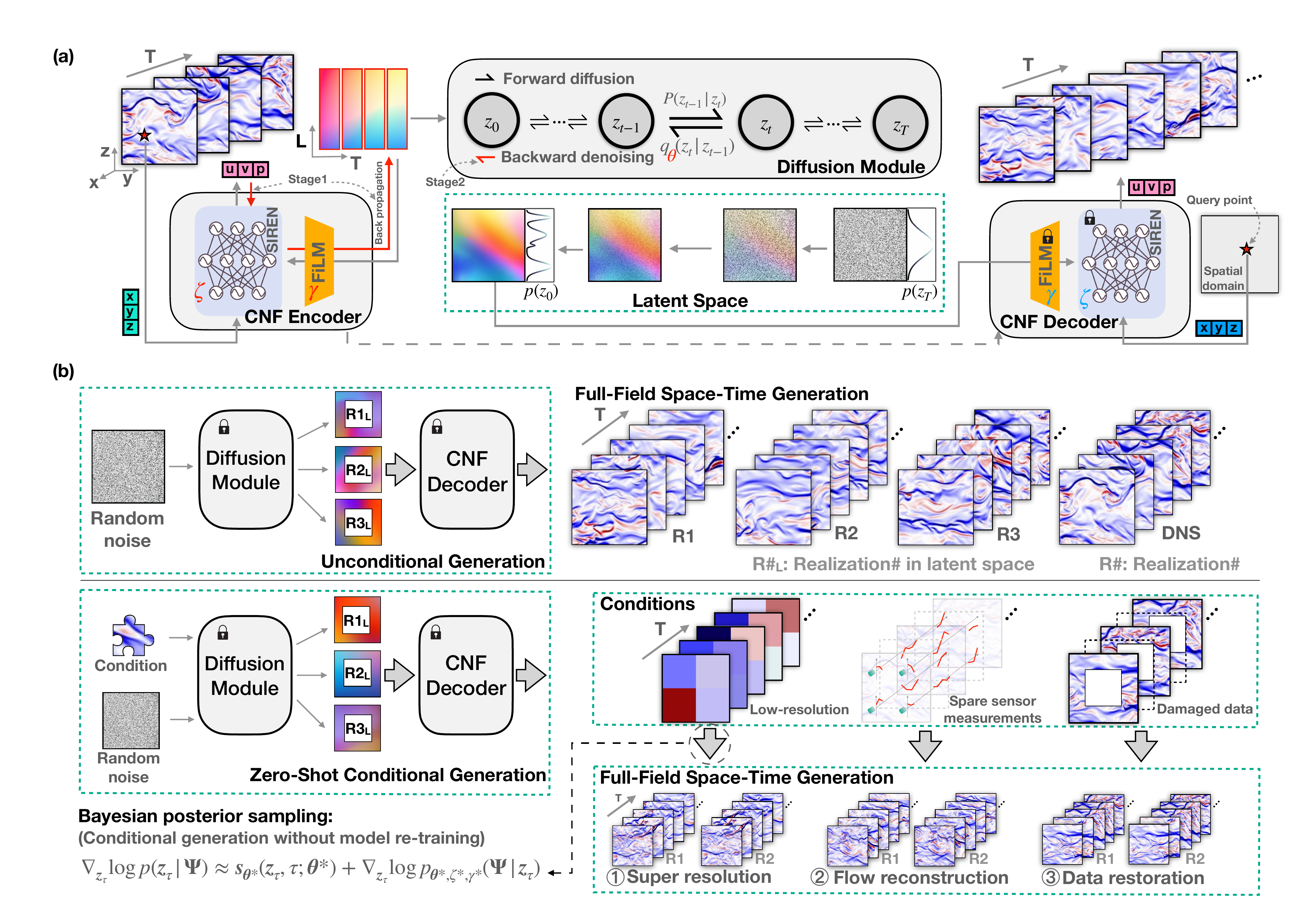}
    \caption{Overview of the proposed Conditional Neural Field Latent Diffusion (CoNFiLD) model. (a) Architectures: a FiLM-based CNF for encoding dynamic flow sequences into latent space, where the underlying distribution of the latent vectors is implicitly captured by learning reverse diffusion (denoising) processes. (b) Zero-shot generation: synthesizing new spatiotemporal flow fields with arbitrary length, either unconditionally or based on specific conditions (e.g., sparse sensor data), without the need for retraining.}  %We first train the CNF encoder to encode each frame of the flow field implicitly to a latent vector (stage 1). Then sequences of latent vectors are passed into the diffusion module for it to learn the intrinsic probability distribution of the spatiotemporal flow field in latent space (stage 2). After trained, the diffusion module generates unseen latent sequence, which can be decoded back to physical space via the CNF decoder. Panel (b): The trained CoNFiLD can generate arbitrarily long new spatiotemporal flow fields unconditionally (first row). Moreover, given different forms of conditions, the CoNFiLD can generate flow fields based on observations with out any re-training, which can be used for various practical applications.}
    \label{fig:method}
\end{figure}

Turbulent flows, inherently chaotic and stochastic across various spatial and temporal scales, fundamentally represent stochastic spatiotemporal processes. The work aims to construct a data-driven model capable of generating unsteady instantaneous turbulent flows. This is achieved through generative AI techniques designed to learn the underlying probability distribution $p\big(\bm{\Phi}(\bm{x}, t)\big)$ of the spatiotemporal turbulent flow fields $\bm{\Phi}(\bm{x}, t)$ from instantaneous flow datasets $\bm{\Phi}(\bm{x}, t) \in \mathcal{A}_{train}$. To this end, we present the Conditional Neural Field Latent Diffusion (CoNFiLD) model, an innovative generative learning framework that leverages neural implicit representations to facilitate efficient and scalable diffusion-based generation within a compact latent space. As depicted in Fig.~\ref{fig:method}, our CoNFiLD features a unique combination of Conditional Neural Field (CNF) and Latent Diffusion Models (LDM), distinguishing it from prior work that applied diffusion processes directly in the high-dimensional physical domain, which encounters significant computational hurdles and memory constraints~\cite{gao2023bayesian}. Specifically, the CoNFiLD model is constructed in three stages. 

First, a CNF, $\mathscr{E}_{\zeta,\gamma}(\mathbf{x}, \bm{L})$, is designed to encode a time sequence of instantaneous flow fields, discretized as $\Phi(\bm{x}, t) \in \mathbb{R}^{N_{m}\times N_{t}}$, into a time sequence of latents $\bm{z}_0 \in \mathbb{R}^{N_l\times N_t}$, where $\zeta,\gamma$ are trainable parameters of the CNF encoder, and $N_{m}, N_{l}, N_{t}$ represent the dimensions of spatial space, latent space, and time length, respectively. Once trained, the CNF forms a neural implicit representation of the spatiotemporal flow field conditioned on the latent vector $\bm{L} = \bm{z}_0$, i.e., $\bm{\Phi}(\bm{x}, t) \approx \mathscr{E}_{\zeta^*,\gamma^*}(\mathbf{x}, \bm{L})$. Unlike conventional encoders, the CNF encoder here is formulated in an auto-decoding fashion~\cite{park2019deepsdf}, where the latents $\bm{L}$ are optimized by minimizing the mismatch between the field of interest values and corresponding CNF outputs,
\begin{equation}
\label{eq:encoding}
\begin{aligned}
\bm{z}_0({:}, t_i) &= \mathrm{Encode}\Big(\bm{\Phi}({:}, t_i), \mathscr{E}_{\zeta^*,\gamma^*}\Big)\\ 
 & = \arg \min_{\bm{L}}\sum_j^{N_m}\Big\|\bm{\Phi}(t_i,\mathbf{x}_j) - \mathscr{E}\left(\bm{L}(i), \mathbf{x}_j;\zeta,\gamma\right)\Big\|_{L_2}, \;\ \forall i \in [1,2,\dots,N_t] 
\end{aligned} 
\end{equation}
In practice, all the snapshots $\{\bm{\Phi}({:}, t_i)\}, i \in [1,\dots,N_t]$ are encoded into latent vectors $\{\bm{z}_0({:}, t_i)\}_1^{N_t}$ simultaneously, forming a latent-time snapshot $\bm{z}_0$ as a 2D ``image".

Following the encoding phase, a probabilistic diffusion module is introduced to implicitly learn the underlying probability distribution $p(\bm{z}_0)$ of the latent dynamics $\bm{z}_0$ through bi-direction diffusion processes. Initially, the latent samples undergo a forward Markovian diffusion process, characterized by a series of carefully designed white noise additions that incrementally nudge the latent representations towards the fully perturbed state with an isotropic Gaussian distribution. Subsequently, by learning the reverse diffusion process (i.e., denoising process) through neural networks, the model is capable of generating new latent samples $\bm{z}_0$ from randomly sampled white noises using the learned denoising scheme. 

Finally, the newly generated latents $\bm{z}_0 \in \mathcal{L}_{test}$ is fed into the trained CNF to decode them back to the physical space for obtaining the synthesized spatiotemporal flow fields $\bm{\Phi}(\bm{x}, t) \in \mathcal{A}_{test}$ (see Fig.~\ref{fig:method}),
\begin{equation}
\begin{aligned}
\bm{\Phi}(t_i, {:}) &= \mathrm{Decode}\Big(\bm{z}_0({:}, t_i), \mathscr{E}_{\zeta^*,\gamma^*}\Big)\\ 
 & = \mathscr{E}\Big(\bm{x}, \bm{z}_0({:}, t_i), ;\zeta^*,\gamma^*\Big), \;\;\;\;\; \forall i \in [1,2,\dots,N_t] 
\end{aligned} 
\end{equation}
The much higher data compression ratios of the CNF-based encoder, compared to other encoding methods, allow the generative model to operate in a significantly reduced-dimensional latent space, which addresses the computational challenges in synthesizing large-scale, high-dimensional spatiotemporal turbulence data. Moreover, the CNF's ability to process arbitrary point queries significantly enhances the model's versatility in managing irregular domains and adaptive meshes. The training of the CoNFiLD model unfolds in a decoupled two-step strategy: firstly, the CNF encoder is trained to transform spatiotemporal flow fields into latent representations; this is followed by the diffusion model being trained on these representations. This dual-phase training strategy facilitates efficient utilization of the latent space and enables robust model optimization and inference.         

Upon completing its training, the CoNFiLD model can rapidly generate new 4D spatiotemporal flow samples, $\bm{\Phi} \in \mathbb{R}^{N_m \times N_T}$. Notably, the length $N_T$ of generated time sequences can significantly exceed the length $N_t$ of those used during training ($N_T > N_t$). This novel feature stems from the shift-invariance of convolution kernels learned in the latent space, where the latent diffusion model is capable of synthesizing an arbitrarily extended sequence of latent vectors, which are subsequently decoded into high-resolution 4D flow fields via the trained CNF decoder. Another distinctive feature of CoNFiLD is its \emph{zero-shot} conditioned generation capability, which enables the creation of 4D flow realizations under specific conditions (e.g., sparse sensor measurements, low-resolution data) \emph{without} the need for retraining the model. Unlike traditional conditional generative methods, which require conditionally paired training data and necessitate retraining for new conditions, CoNFiLD's diffusion process is trained unconditionally only once and can then generate samples under a variety of conditions during inference. As shown in Fig.~\ref{fig:method}, this novel feature holds significant practical value, finding applications in a range of inverse problems, such as spatiotemporal super-resolution of flow data (super-resolution), full-field reconstruction of instantaneous flow fields from sparse sensor measurements (flow reconstruction), and restoring missing information in damaged flow data (data restoration). The major highlights of our method include: (1) CNF-based encoding with a high compression ratio, facilitating efficient diffusion processes within the latent space; (2) the ability of CNF to process arbitrary pointwise queries, enhancing adaptability to irregular domains and enabling support for unstructured data and adaptive meshes; (3) a Bayesian conditioning sampling mechanism that allows for versatile conditional generation without the necessity for retraining; (4) significant reduction of memory usage in subsampling-based conditional generation scenarios.

\subsection{Conditional neural field encoding}
\label{sec:encoding}

Neural fields (NF) have emerged as state-of-the-art in learning implicit representations of coordinate-based functional fields, demonstrating exceptional performance in modeling and compressing complex signals such as images~\cite{sitzmann2020implicit, tancik2020fourier}, videos~\cite{Li_2022_CVPR,chen2021nerv}, 3D scenes~\cite{mildenhall2021nerf}, and 3D shapes~\cite{park2019deepsdf,chen2019learning}. Despite their tremendous success in computer vision, the exploration of NFs in dimension reduction of large-scale spatiotemporal data of physical systems (e.g., turbulence data) remains sparse. A recent effort in this direction by Pan et al.~\cite{pan2023neural} employs a neural field fully conditioned by hyper-network for spatiotemporal dimension reduction, where the parameters of the NF are fully determined by the hyper-network. This architecture, while effective, necessitates explicit conditions (e.g., sensor signals, time stamps) and considerable memory overhead for the hyper-network, often surpassing that of the NF itself. In contrast, the CNF of our CoNFiLD utilizes a more flexible and robust conditioning mechanism, the Feature-wise Linear Modulation (FiLM)~\cite{perez2018film}, and the encoder is formulated in an auto-decoding manner. The primary NF leverages the SIREN network architecture~\cite{sitzmann2020implicit}, renowned for its capacity to capture domains with rich periodic features through sinusoidal activation functions. The mathematical representation is as follows,
\begin{equation}
\begin{gathered}
\mathrm{SIREN}(\bm{x}) = \bm{W}_p\left(\eta_{p-1}\circ\eta_{p-2}\circ\dots\circ\eta_1\right)\left(\omega_0\bm{W}_0\bm{x} + \bm{B}_0\right) + \bm{B}_p,  \\
\eta_i(\bm{o}_{i-1}) = \sin(\bm{W}_i \bm{o}_{i-1} + \bm{B}_i), \;\;\; i \in [1,2,\dots,p],
\end{gathered}
\end{equation}
where $\{\bm{W}_i\}^p_{0}$ and $\{\bm{B}_i\}^p_{0}$ are trainable parameters of NF (collectively denoted as $\zeta$), $\bm{o}_{i-1}$ is the $i-1^\mathrm{th}$ layer output, and $\omega_0$ is a hyperparameter that modulates the initial input signal frequency. SIREN requires a unique initialization, $w_i \sim \mathcal{U}(-r / \sqrt{n},r / \sqrt{n}), r \in \mathbb{R}$, ensure outputs across layers follow a standard normal distribution, where $w_i$ is individual entry of weight matrices and $\mathcal{U}(\cdot)$ represents a uniform distribution with lower/upper bound hyperparameters $r$. Typically, $\omega_0 = 30$ and $r = \sqrt{6}$ are chosen for robust performance.

The SIREN is modulated by FiLM conditioning, where latent vectors $\bm{L}$ are passed into multiple linear transformation layers. Thus, the CNF, denoted by $\mathscr{E}$, is described as follows,
\begin{equation}
\begin{gathered}
\mathscr{E}(\bm{x}, \bm{L}) = \bm{W}_p \left(\eta'_{p-1}\circ\eta'_{p-2}\circ\dots\circ\eta'_1\right)\left(\omega_0\bm{W}_0\bm{x} + \bm{B}_0 + \bm{c}_0\right) + \bm{B}_p,  \\
\eta'_i(\bm{o}_{i-1}, \bm{c}_i) = \sin(\bm{W}_i\bm{o}_{i-1} + \mathbf{B}_i + \bm{c}_i), \;\;\; i \in [1,2,\dots,p], \\ 
\bm{c}_i(\bm{L})= \bm{W}^{cond}_i\bm{L}+\mathbf{B}^{cond}_i,
\end{gathered}
\end{equation}
where $\{\bm{W}^{cond}_i\}^p_{1}$ and $\{\bm{B}^{cond}_i\}^p_{1}$ (collectively denoted as $\gamma$) are trainable parameters for the FiLM conditioning layers, which introduce a bias adjustment $\bm{c}_i$ on each SIREN layer before the activation function is applied and $\eta'_i$ represents the $i^\mathrm{th}$ FiLM-modified intermediate layer. This implicit model represents a continuous function across the spatial domain $\Omega$, usually being trained on discretized datasets. Within our CoNFiLD framework, the CNF functions dually as an encoder and decoder, enabling seamless data transform between physical and latent space. The encoding strategy was meticulously designed to ensure that the SIREN exclusively models the spatial field, whereas the conditional information is encapsulated within a latent vector corresponding to each spatial field frame at discrete time steps $t_i$. Consequently, the latent encoding of a spatiotemporal field segment is structured as a 2D ``image", effectively preserving the original time dimension. This configuration enables unrestricted kernel convolution across images of any temporal length, introducing an innovative approach for time forecasting in the generation process, further detailed in Section~\ref{sec:diffusion}.

The CNF is trained on the dataset $\bm{\Phi} \in \mathcal{A}_{train}$ by solving the following optimization,
\begin{equation}
\begin{aligned}
\bm{L}^*, \zeta^*, \gamma^* = \arg \min_{L,\zeta,\gamma} \sum_i^{N_t}\sum_j^{N_m} \Big\|\bm{\Phi}(\bm{x}_j, t_i) - \mathscr{E}\Big(\bm{L}(i), \bm{x}_j; \zeta,\gamma \Big) \Big\|_{L_2},
\end{aligned} 
\end{equation}
where the optimized latent vectors $\bm{L}^*$, representing the latents $\bm{z}_0$ of the spatiotemporal flow field $\bm{\Phi}$, are obtained. This optimization employs an alternating-direction strategy, which involves updating the CNF parameters ($\zeta, \gamma$) and the latent vectors ($\bm{L}$) in turns. Specifically, the latent vectors are updated per batch with the CNF parameters frozen; subsequently, $\zeta$ and $\gamma$ are updated while the latent vectors remain temporarily fixed. This alternating-direction optimization approach has been empirically shown to foster a stable and robust convergence during the training process~\cite{serrano2024operator}. After training, encoding a new spatiotemporal field necessitates an optimization, as described by Eq.~\ref{eq:encoding}, whereas the decoding phase simply involves feeding the latent vectors and spatial coordinates into the trained CNF to retrieve the corresponding Field of Interest (FOI) values. The efficiency of the CNF decoder becomes particularly evident during the latent diffusion-based generation phase.

\subsection{Latent probabilistic diffusion modeling}
\label{sec:diffusion}

Given a CNF-encoded latent state $\bm{z}_0$ whose underlying distribution is $p(\bm{z}_0)$, i.e., $\bm{z}_0 \sim p(\bm{z}_0)$, a \emph{forward diffusion process} is defined by progressively perturbing $\bm{z}_0$ with Gaussian noise of variance $\beta_\tau$, through the transition kernel,
\begin{equation}
    p(\bm{z}_\tau|\bm{z}_{\tau-1}) = \mathcal{N}(\bm{z}_\tau; \sqrt{1-\beta_\tau}\bm{z}_{\tau-1}, \beta_\tau\bm{I}),
\end{equation}
where $\tau = 1, \cdots N_\tau$ denotes the diffusion step index with $N_\tau$ as the total number of steps, and $\bm{I} \in \mathbb{R}^{(N_l \times N_t)^2}$ is the identity matrix. This forward diffusion process yields a sequence of incrementally noised latent states $\bm{z}_1, \cdots, \bm{z}_{N_\tau}$ with the joint probability density $p(\bm{z}_1, \cdots, \bm{z}_{N_\tau}|\bm{z}_0)$,
\begin{equation}
    p(\bm{z}_1, \cdots, \bm{z}_{N_\tau}|\bm{z}_0) = \prod_{\tau=1}^{N_\tau}q(\bm{z}_\tau|\bm{z}_{\tau-1}). 
\end{equation}
With sufficient perturbation steps, the marginalized distribution $p(\bm{z}_{N_\tau}|\bm{z}_0)$ asymptotically converges to an isotropic Gaussian distribution, denoted by $p(\bm{z}_\tau) = \mathcal{N}(\bm{0}, \sigma^2_{N_\tau|0}\bm{I})$, facilitating straightforward sampling. Using the re-parameterization trick~\cite{ho2020denoising}, the conditional distribution of each noised latent state given $\bm{z}_0$ is also Gaussian, explicitly defined as,
\begin{equation}
    p(\bm{z}_\tau|\bm{z}_0) = \mathcal{N}(\bm{z}_\tau, \sqrt{\bar{\alpha}_\tau}\bm{x}_0, (1-\bar{\alpha}_\tau)
    \bm{I})), 
\end{equation}
where $\alpha_\tau = 1 - \beta_\tau$ and $\bar{\alpha}_\tau = \prod_{s=1}^\tau \alpha_s$. The forward diffusion is characterized through a predetermined series of variance parameters $\beta_\tau$, known as variance schedule, which can adopt various forms, such as linear, quadratic, or cosine schedules~\cite{yang2023diffusion}.   

Upon establishing the forward diffusion process, its reversal (i.e., \emph{reverse diffusion process}) becomes particularly compelling since it enables the synthesis of new latent samples of $\bm{z}_0$ from white noise vectors sampled from the isotropic Gaussian distribution $p(\bm{z}_{N_\tau})$. This process relies on the assumption that each step's perturbation is sufficiently small, ensuring that the conditional probability $p(\bm{z}_{\tau-1}|\bm{z}_\tau)$, or the reverse transition kernel, remains Gaussian. However, directly computing this reverse transition kernel is infeasible, as it requires the knowledge of the distribution of the entire latent space, which is unknown \emph{a priori}. To overcome this challenge, all existing diffusion-based generation methods rely on neural networks to learn either the reverse transition kernel or the score function with trainable parameters $\bm{\theta}$, both of which can be unified in the same framework~\cite{song2019generative,ho2020denoising,song2020score}. Training the neural network parameterization can be formulated as a likelihood maximization problem,
\begin{equation}
    \min_{\bm{\theta}}\sum_{\mathscr{E}_{\zeta^*, \gamma^*}(\bm{z}_0) \in \mathcal{A}_{train}} {-\log p_{\bm{\theta}}(\bm{z}_0)}. 
\end{equation}
However, as $\log p_{\bm{\theta}}(\bm{z}_0)$ is not tractable, we can instead minimize the variational bound $L_\mathrm{VB}$ on the negative log likelihood (derivation can be found in \cite{sohl2015deep}),
\begin{equation}
\begin{aligned}
    \mathbb{E}\big[-\log p_{\bm{\theta}}(\bm{z}_0)\big] \leq L_\mathrm{VB} =\ & \mathbb{E}_{p}\bigg[D_{KL}\Big(p(\bm{z}_{N_\tau}|\bm{z}_0) \big\| p(\bm{z}_{N_\tau})\Big)\\
    + & \sum_{\tau=2}^{N_\tau}D_{KL}\Big(p(\bm{z}_{\tau-1}|\bm{z}_{\tau}, \bm{z}_{0}) \big\| p_{\bm{\theta}}(\bm{z}_{\tau-1}|\bm{z}_\tau)\Big) - \log p_{\bm{\theta}}(\bm{z}_0|\bm{z}_1)\bigg],
\end{aligned}
\label{eq:VB}
\end{equation}
where $D_{KL}(\cdot\|\cdot)$ represents Kullback-Leibler (KL) divergence operator. Since the first term of $L_\mathrm{VB}$ contains no trainable parameters $\bm{\theta}$ and remains a constant during training, it can be dropped. Therefore, the diffusion model can be trained by minimizing $\tilde{L}_\mathrm{VB}$,
\begin{equation}
    \tilde{L}_\mathrm{VB} = \mathbb{E}_{p}\bigg[\sum_{\tau=2}^{N_\tau}D_{KL}\Big(p(\bm{z}_{\tau-1}|\bm{z}_{\tau}, \bm{z}_{0}) \big\| p_{\bm{\theta}}(\bm{z}_{\tau-1}|\bm{z}_\tau)\Big) - \log p_{\bm{\theta}}(\bm{z}_0|\bm{z}_1)\bigg].
\label{eq:vb}
\end{equation}
The reverse transition kernel is Gaussian, $p_{\bm{\theta}}(\bm{z}_{\tau-1}|\bm{z}_\tau, \bm{z}_0) = \mathcal{N}\Big(\bm{z}_{\tau-1}; \bm{\mu}_{\bm{\theta}}(\bm{z}_\tau, t), \bm{\Sigma}_{\bm{\theta}}(\bm{z}_\tau, t)\Big)$, where the mean and covariance functions are parameterized by neural networks. Specifically, the parameterization is designed based on the reverse conditional probability $p(\bm{z}_{\tau-1}|\bm{z}_{\tau}, \bm{z}_0)$, which is analytically tractable when conditioned on $\bm{z}_0$,
\begin{equation}
    p(\bm{z}_{\tau-1}|\bm{z}_{\tau}, \bm{z}_0) = \mathcal{N}\bigg(\bm{z}_{\tau-1}; \frac{1}{\sqrt{\alpha_\tau}}\left(\bm{x}_\tau - \frac{1-\alpha_\tau}{\sqrt{1-\bar{\alpha}_\tau}}\bm{\epsilon}_\tau\right), \frac{1-\bar{\alpha}_{\tau-1}}{1-\bar{\alpha}_\tau}\beta_\tau \bm{I} \bigg)
    \label{eq:reverse-kernel}
\end{equation}
where the $\bm{\epsilon}_\tau$ is the noise added at step $\tau$. Therefore, the mean function $\bm{\mu}_{\bm{\theta}}(\bm{z}_\tau, \tau)$ of reverse kernel is parameterized as follows,
\begin{equation}
    \bm{\mu}_{\bm{\theta}}(\bm{z}_\tau, \tau) = \frac{1}{\sqrt{\alpha_\tau}}\left(\bm{x}_\tau - \frac{1-\alpha_\tau}{\sqrt{1-\bar{\alpha}_\tau}}\bm{\epsilon}_{\bm{\theta}}(\bm{z}_\tau, \tau; \bm{\theta})\right)
\end{equation}
where the noise function $\bm{\epsilon}_{\bm{\theta}}(\bm{z}_\tau, \tau; \bm{\theta})$ is approximated by a U-Net variant with residual blocks, self-attention blocks, and group normalization~\cite{nichol2021improved,dhariwal2021diffusion}. The variance function $\bm{\Sigma}_\theta$ remains fixed as $\bm{\Sigma}_\theta = (1-\bar{\alpha}_{\tau-1})\beta_\tau/(1-\bar{\alpha}_\tau) \bm{I}$ based on Eq.~\ref{eq:reverse-kernel}. Since the KL divergence between two Gaussian distributions has a closed form, $\tilde{L}_\mathrm{VB}$ in Eq.~\ref{eq:vb} can be expressed as,
\begin{equation}
    \tilde{L}_\mathrm{VB} = \mathbb{E}_{\bm{z}_0, \bm{\epsilon}}\bigg[
    \frac{(1-\alpha_\tau)^2}{2\alpha_\tau(1-\bar{\alpha}_\tau)\|\bm{\Sigma}_{\bm{\theta}}\|^2_{L_2}} \Big\| \bm{\epsilon}_\tau - \bm{\epsilon}_{\bm{\theta}}\big(\sqrt{\bar{\alpha}_t} \bm{z}_0 + \sqrt{1-\bar{\alpha}_t}\bm{\epsilon}_\tau, \tau \big) \Big\|^2_{L_2}
    \bigg].
\end{equation}
Ho et al.~\cite{ho2020denoising} further simplified the VB loss by ignoring the weighting term,
\begin{equation}
    \tilde{L}_\mathrm{simple} = \mathbb{E}_{\bm{z}_0, \bm{\epsilon}}\bigg[\Big\| \bm{\epsilon}_\tau - \bm{\epsilon}_{\bm{\theta}}\big(\sqrt{\bar{\alpha}_t} \bm{z}_0 + \sqrt{1-\bar{\alpha}_t}\bm{\epsilon}_\tau, \tau \big) \Big\|^2_{L_2}
    \bigg].
\end{equation}
In this work, the latent diffusion model training adopts a hybrid form suggested by Nichol and Dhariwal~\cite{nichol2021improved}, which uses both $\tilde{L}_\mathrm{simple}$ and $\tilde{L}_\mathrm{vb}$ with a weight parameter $\lambda$, leading to the following optimization,
\begin{equation}
    \bm{\theta}^* = \arg \min_{\bm{\theta}} \sum_{\mathscr{E}_{ \zeta^*, \gamma^*}(\bm{z}_0) \in \mathcal{A}_{train}} \Big[\tilde{L}_\mathrm{simple} + \lambda \tilde{L}_\mathrm{VB}\Big], 
\end{equation}
where noise vector $\bm{\epsilon}$ is randomly sampled from a standard Gaussian distribution, i.e., $\bm{\epsilon} \sim \mathcal{N}(\bm{0}, \bm{I})$, and $\tau$ is randomly sampled from a discrete uniform distribution $\mathcal{U}(1, N_\tau)$.  

\subsection{Bayesian conditional generation of spatiotemporal fields}
\label{sec:uncon generation}
Upon completing the training phase, the CoNFiLD model can rapidly generate new 4D spatiotemporal flow fields $\bm{\Phi} \in \mathbb{R}^{N_m \times N_T}$ by sampling from the latent diffusion model and performing the CNF decoding. Remarkably, the model enables the generation of time sequences of flow fields that far exceed the temporal scope of the training data, i.e., $N_T > N_t$. This extended generative capacity allows the CoNFiLD to not only synthesize new spatiotemporal flow data possessing turbulence statistics consistent with the training data, but also to extrapolate temporally well beyond the time sequence length of the training set. More importantly, a particularly notable feature of the CoNFiLD model is its adeptness at \emph{zero-shot} conditional generation within a Bayesian sampling framework. This capability allows for the spatiotemporal generation under various conditions, such as specific initial states, sparse observations, or low-fidelity simulations, all without necessitating retraining for each unique scenario. This aspect highlights the CoNFiLD's versatility and adaptability, offering tailored turbulence predictions based on available data or desired outcomes. 

\subsubsection{Unconditional generation}
The trained CoNFiLD model can be utilized to generate new spatiotemporal turbulent flow sequences by sampling the learned distribution via the reverse diffusion process. Starting from samples of a multivariate isotropic Gaussian distribution, $\bm{z}_{N_\tau} \sim \mathcal{N}(\bm{0}, \bm{I})$, the noises are progressively removed using the learned reverse transition kernel $p_{\bm{\theta}^*}(\bm{z}_{\tau-1}|\bm{z}_{\tau}, \bm{z}_0)$, 
\begin{equation}
    p_{\bm{\theta}^*}(\bm{z}_{\tau-1}|\bm{z}_{\tau}, \bm{z}_0) = \mathcal{N}\bigg(\bm{z}_{\tau-1}; \frac{1}{\sqrt{\alpha_\tau}}\left(\bm{z}_\tau - \frac{1-\alpha_\tau}{\sqrt{1-\bar{\alpha}_\tau}}\bm{\epsilon}_{\bm{\theta}^*}(\bm{z}_\tau, \tau; \bm{\theta}^*)\right), \frac{1-\bar{\alpha}_{\tau-1}}{1-\bar{\alpha}_\tau}\beta_\tau \bm{I} \bigg),
    \label{eq:reverse-kernel-denoise}
\end{equation}
allowing for the sampling of $\bm{z}_{\tau-1}$ based on the sample $\bm{z}_{\tau}$ at step $\tau$. Namely, the following reverse sampling is conducted,
\begin{equation}
\bm{z}_{\tau-1} = \frac{1}{\sqrt{\alpha_\tau}}\left(\bm{z}_\tau - \frac{1-\alpha_\tau}{\sqrt{1-\bar{\alpha}_\tau}}\bm{\epsilon}_{\bm{\theta}^*}(\bm{z}_\tau, \tau; \bm{\theta}^*)\right) + \sigma_\tau\bm{\epsilon}, \ \ \bm{\epsilon}\sim\mathcal{N}(\bm{0}, \bm{I}).
\label{eq:UG}
\end{equation}
By iteratively applying Eq.~\ref{eq:UG}, a new latent $\bm{z}_0$ starting from white noise $\bm{z}_{N_\tau}$ can be obtained, which is then decoded back to a spatiotemporal turbulent flow field $\bm{\Phi} = \mathcal{E}_{\zeta^*,\gamma^*}(\bm{x}, \bm{z}_0)$ via the CNF decoder. This process aligns with the score-based generative modeling framework~\cite{song2019generative}, where Eq.~\ref{eq:UG} can be expressed in terms of the score function,
\begin{equation}
\bm{z}_{\tau-1} = \frac{1}{\sqrt{\alpha_\tau}}\Big(\bm{z}_\tau + (1-\alpha_\tau)\bm{s}_{\bm{\theta}^*}(\bm{z}_\tau, \tau; \bm{\theta}^*)\Big) + \sigma_\tau\bm{\epsilon}, \ \ \bm{\epsilon}\sim\mathcal{N}(\bm{0}, \bm{I}).
\label{eq:UG-score}
\end{equation}
where the Stein score function $\bm{s}(\bm{z}_\tau, \tau) = \nabla_{\bm{z}_t}\log p(\bm{z}_t)$ is approximated by neural network parameterization $\bm{s}_{\bm{\theta}^*}(\bm{z}_\tau, \tau; \bm{\theta}^*)$ with optimized parameters $\bm{\theta}^*$.

\subsubsection{Conditional generation}
Mathematically, conditions can be systematically represented by a vector $\bm{\Psi} \in \mathbb{R}^{N_{\bm{\Psi}}}$, hereafter referred to as the condition vector. This vector $\bm{\Psi}$ may represent low-fidelity solutions, instantaneous flow measurements from sparse sensor array, low-resolution observation data, or other related information regarding to the spatiotemporal flow field of interest $\bm{\Phi}$. From Bayesian perspective, conditional generation involves the sampling of the conditional probability $p(\bm{\Phi}|\bm{\Psi})$, which can be bridged to the unconditioned density $p(\bm{\Phi})$ via Bayes' rule,
\begin{equation}
    p(\bm{\Phi}|\bm{\Psi}) \propto p(\bm{\Psi}|\bm{\Phi})p(\bm{\Phi}), 
\end{equation}
where $p(\bm{\Phi})$ is learned by the CoNFiLD after the unconditional training,
\begin{equation}
    p(\bm{\Phi}) \approx p^\mathrm{CoNFiLD}(\bm{\Phi};\bm{\theta}^*,\zeta^*,\gamma^*). 
\end{equation}   
The relationship between the condition $\bm{\Psi}$ and corresponding flow sequences $\bm{\Phi}$ is defined as follows,
\begin{equation}
    \bm{\Psi} = \mathcal{F}(\bm{\Phi}) + \bm{\epsilon}_c,
    \label{eq:nonlinearmap}
\end{equation} 
where $\mathcal{F}: \mathbb{R}^{N_m \times N_T} \to \mathbb{R}^{N_{\bm{\Psi}}}$ is a nonlinear mapping from the spatiotemporal turbulent field $\bm{\Phi}$ to its associated condition vector $\bm{\Psi}$ (e.g., partial observation); $\bm{\epsilon}_c$ represents the uncertainty inherent in the state-to-condition mapping, typically modeled as a zero-mean Gaussian random variable. Conditional generation can be conceptualized as a Bayesian inverse problem, which has attracted significant interest within the computer vision community, notably in the area of image restoration~\cite{li2023diffusion}. In these contexts, the state-to-condition mapping often exhibits a linear nature. The Diffusion Posterior Sampling (DPS) technique, which accommodates both linear and nonlinear mappings within the pixel space, represents a recent advancement in this area, albeit operating within the pixel space~\cite{chung2022diffusion}. Expanding upon the DPS concept, we adapt it to the CNF-encoded latent diffusion process, facilitating \emph{zero-shot} conditional generation. In CoNFiLD, Eq.~\ref{eq:nonlinearmap} is reformulated as,
\begin{equation}
    \bm{\Psi} = \mathcal{F}\Big(\mathcal{E}_{\zeta^*,\gamma^*}(\bm{z}_0; \zeta^*,\gamma^*)\Big) + \bm{\epsilon}_c.
    \label{eq:latenttoobservation}
\end{equation} 
The conditional relationship between the perturbed latent states and the conditions $\bm{\Psi}$ is expressed via Bayes's theorem as $p(\bm{z}_\tau|\bm{\Psi}) = {p(\bm{\Psi}|\bm{z}_\tau) p(\bm{z}_\tau)}/p(\bm{\Psi})$, with the normalizing constant $p(\bm{\Psi})$ being generally intractable. This complexity is circumvented by shifting the formulation to score functions,
\begin{equation}
    \nabla_{\bm{z}_\tau}\log p(\bm{z}_\tau|\bm{\Psi}) = \nabla_{\bm{z}_\tau}\log p(\bm{\Psi}|\bm{z}_\tau) + \nabla_{\bm{z}_\tau} \log p(\bm{z}_\tau),
    \label{eq:con-score}
\end{equation} 
where second term $\nabla_{\bm{z}_\tau} \log p(\bm{z}_\tau)$ is the score function for unconditional generation, approximated by $\bm{s}_{\bm{\theta}^*}(\bm{z}_\tau, \tau)$ post-training. However, the gradient of log-likelihood term, $\nabla_{\bm{z}_\tau}\log p(\bm{\Psi}|\bm{z}_\tau)$, remains to be estimated. To this end, the likelihood density is factorized,
\begin{equation}
\begin{aligned}
    p(\bm{\Psi|\bm{z}_\tau}) &= \int p(\bm{\Psi}|\bm{z}_0, \bm{z}_\tau)p(\bm{z}_0|\bm{z}_\tau)d\bm{z}_0
    = \int p(\bm{\Psi}|\bm{z}_0)p(\bm{z}_0|\bm{z}_\tau)d\bm{z}_0 \\
    &= \mathbb{E}_{\bm{z}_0 \sim p(\bm{z}_0|\bm{z}_\tau)}\Big[p(\bm{\Psi}|\bm{z}_0)\Big],
\end{aligned}
\end{equation}
which can be approximated by,
\begin{equation}
    p(\bm{\Psi}|\bm{z}_\tau) = \mathbb{E}_{\bm{z}_0 \sim p(\bm{z}_0|\bm{z}_\tau)}\big[p(\bm{\Psi}|\bm{z}_0)\big] \approx p\big(\bm{\Psi}|\mathbb{E}[\bm{z}_0|\bm{z}_\tau]\big).
\end{equation}
The approximation error is theoretically bounded bounded with the Jensen gap~\cite{gao2017bounds}. Accordingly, the gradient of the log likelihood can be approximated as,
\begin{equation}
    \nabla_{\bm{z}_\tau} \log p(\bm{\Psi}|\bm{z}_\tau) \approx  \nabla_{\bm{z}_\tau} \log p\big(\bm{\Psi}|\mathbb{E}[\bm{z}_0|\bm{z}_\tau]\big)
\end{equation}
where the posterior mean $\hat{\bm{z}}_0 = \mathbb{E}[\bm{z}_0|\bm{z}_\tau]$ can be computed as,
\begin{equation}
    \hat{\bm{z}}_0 = \mathbb{E}[\bm{z}_0|\bm{z}_\tau] = \frac{1}{\sqrt{\bar{\alpha}_\tau}}\bigg(\bm{z}_\tau + (1-\bar{\alpha}_\tau) \nabla_{\bm{z}_\tau} \log p(\bm{z}_\tau) \bigg),
\end{equation}
where the Stein score $\nabla_{\bm{z}_\tau} \log p(\bm{z}_\tau)$ has been learned during the unconditional diffusion modeling training, i.e., $\nabla_{\bm{z}_\tau} \log p(\bm{z}_\tau) \approx \bm{s}_{\bm{\theta}^*}(\bm{z}_\tau, \tau; \bm{\theta}^*)$. The we have approximated $\hat{\bm{z}}_0$,
\begin{equation}
    \hat{\bm{z}}_0 \approx \hat{\bm{z}}_0^*(\bm{z}_\tau, \tau; \bm{\theta}^*)  = \frac{1}{\sqrt{\bar{\alpha}_\tau}}\bigg(\bm{z}_\tau + (1-\bar{\alpha}_\tau) \bm{s}_{\bm{\theta}^*}(\bm{z}_\tau, \tau; \bm{\theta}^*) \bigg),
\end{equation}
Based on the nested probabilistic nonlinear relationship between the condition $\bm{\Psi}$ and CNF-encoded flow field latents $\bm{z}_0$, the approximated likelihood is,
\begin{equation}
    p(\bm{\Psi}|\bm{z}_0) \approx p(\bm{\Psi}|\bm{z}_0^*) \sim \mathcal{N}\Bigg(\mathcal{F}\bigg(\mathcal{E}_{\zeta^*,\gamma^*}\Big(\bm{z}_0^*(\bm{z}_\tau, \tau; \bm{\theta}^*); \zeta^*,\gamma^*\Big) \bigg), \sigma_c^2\bm{I}\Bigg),
\end{equation}
if the uncertainty term $\epsilon_c$ in Eq.~\ref{eq:nonlinearmap} has a zero-mean Gaussian distribution, i.e., $\epsilon_c \sim \mathcal{N}(\bm{0}, \sigma_c^2\bm{I})$. By differentiating the approximated log likelihood with respect to $\bm{z}_\tau$, 
\begin{equation}
    \nabla_{\bm{z}_\tau}\log p(\bm{z}_\tau|\bm{\Psi}) \approx -\frac{1}{\sigma_c^2}\nabla_{\bm{z}_\tau} \Bigg\| \bm{\Psi} - \mathcal{F}\bigg(\mathcal{E}_{\zeta^*,\gamma^*}\Big(\bm{z}_0^*(\bm{z}_\tau, \tau; \bm{\theta}^*); \zeta^*,\gamma^*\Big) \bigg)  \Bigg\|^2_{L2},
\end{equation}
which can be computed using the chain rule,
\begin{equation}
\begin{aligned}
    \nabla_{\bm{z}_\tau}\log p(\bm{z}_\tau|\bm{\Psi}) &\approx \nabla_{\bm{z}_\tau}\log p_{\bm{\theta}^*, \zeta^*, \gamma^*}(\bm{\Psi}|\bm{z}_\tau) \\
    &= -\frac{2}{\sigma_c^2}(\bm{\Psi} - \mathcal{F}(\mathcal{E}_{\zeta^*, \gamma^*}))\frac{\partial \mathcal{F}(\mathcal{E}_{\zeta^*, \gamma^*})}{\partial \mathcal{E}_{\zeta^*, \gamma^*}} \frac{\partial \mathcal{E}_{\zeta^*, \gamma^*}(\bm{z}_0^*)}{\partial \bm{z}_0^*} \frac{\partial \bm{z}_0^*(\bm{z}_\tau, \tau; \bm{\theta}^*)}{\partial \bm{z}_\tau},
\end{aligned}
\end{equation}
by leveraging the automatic differentiation (AD) capability using differentiable programming for the implementation. Finally, the gradient of log posterior (i.e., guided score function $\bm{s}^\mathrm{guided}_{\bm{\theta}^*,\zeta^*, \gamma^*}$) as shown by Eq.~\ref{eq:con-score}, can be computed as,
\begin{equation}
\begin{aligned}
    \nabla_{\bm{z}_\tau}\log p(\bm{z}_\tau|\bm{\Psi}) 
    &\approx \nabla_{\bm{z}_\tau}\log p_{\bm{\theta}^*, \zeta^*, \gamma^*}(\bm{\Psi}|\bm{z}_\tau) + \nabla_{\bm{z}_\tau} \log p_{\bm{\theta}^*, \zeta^*, \gamma^*}(\bm{z}_\tau) \\
    & = p_{\bm{\theta}^*, \zeta^*, \gamma^*}(\bm{\Psi}|\bm{z}_\tau) + \bm{s}_{\bm{\theta}^*}(\bm{z}_\tau, \tau; \bm{\theta}^*) \\
    & = \bm{s}^\mathrm{guided}_{\bm{\theta}^*,\zeta^*, \gamma^*}(\bm{\Psi}, \bm{z}_\tau, \tau; \zeta^*, \gamma^*, \bm{\theta}^*).
\label{eq:con-score-comput}
\end{aligned}
\end{equation} 
Therefore, conditional sampling can be achieved by modifying the unconditional score function as above. Without the need of retraining the CoNFiLD, new spatiotemporal turbulent flow fields $\bm{\Phi}|\bm{\Psi}$ given conditions $\bm{\Psi}$ can be rapidly generated using the guided score function $\bm{s}^\mathrm{guided}_{\bm{\theta}^*,\zeta^*, \gamma^*}$ derived in Eq.~\ref{eq:con-score-comput}.

\section{Numerical Results}
\label{sec:result}
In this section, we conduct extensive numerical experiments to assess the performance of our proposed CoNFiLD method on a variety of stochastic spatiotemporal flow generation scenarios, including irregular pipe flow with stochastic forcing, turbulent channel flow, flow over periodic hills, and wall-bounded turbulence with roughness, highlighting the model's proficiency in navigating both regular and irregular geometries and managing scenarios with varying flow separation. The dynamics of these fluid flows are governed by the unsteady incompressible Navier-Stokes (NS) equations. 
\begin{equation}
\begin{split}
    \frac{\partial \bm{u}}{\partial t} + (\bm{u} \cdot \nabla)\bm{u} &= -\nabla p + \nu\nabla^2\bm{u} + \bm{f}, \\
    \nabla\cdot\bm{u} &= 0,
    \label{eq:ns}
\end{split}
\end{equation}
where $\bm{u}(\bm{x}, t)$ denotes the velocity vector, $p(\bm{x}, t)$ the pressure, $\nu$ the viscosity, and $\bm{f}(\bm{x},t)$ the forcing term. We will first present the model's capability to synthesize new 4D instantaneous flow fields across these scenarios, with a comparison against DNS references. Additionally, the trained CoNFiLD will be used for zero-shot conditional generation for various data assimilation and inverse problem applications without the need for retraining. These applications range from the full-field reconstruction of flow sequences from sparse sensor measurements to super-resolved spatiotemporal generation and turbulence data restoration.            

\subsection{Unconditional generation of spatiotemporal flow fields}
\subsubsection{Two dimensional irregular pipe flow with stochastic forcing term}
\begin{figure}[t!]
    \centering
    \includegraphics[width=\textwidth]{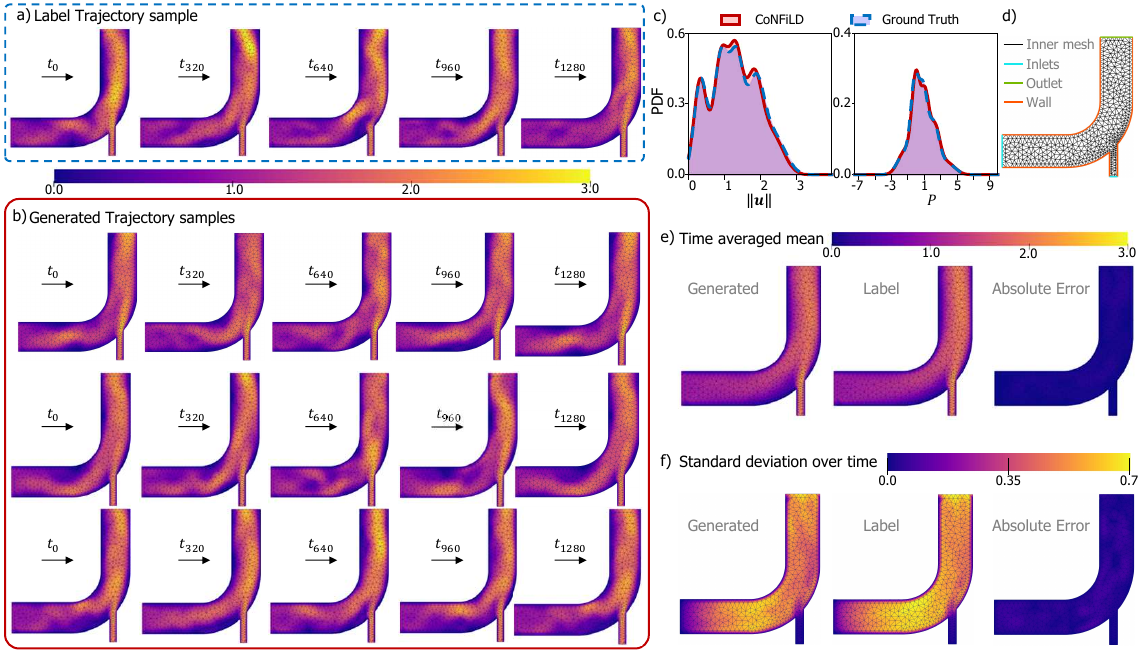}
    \caption{Unconditional generation of flow in a 2D irregular pipe with stochastic forcing. (a) A trajectory of velocity magnitude ($\|\bm{u}\|$) fields of the DNS data $\bm{\Phi}^{dns}$ (ground truth). (b) Three randomly generated flow sequence samples by CoNFiLD (velocity magnitude fields at selected time steps). (c) Comparison of the PDF of the velocity magnitude (left panel) and pressure $p$ (right panel) between the CoNFiLD generated samples and the DNS labels (ground truth). (d) The irregular computational domain with unstructured grids. (e-f) The comparison of the time-averaged mean (e) and standard deviation over time (f) between the generated samples (left) and label data (middle), with the absolute discrepancy (right).}
    \label{fig:elbow}
\end{figure}

We begin with a 2D flow within an irregular pipe subject to stochastic forcing to demonstrate CoNFiLD's capability of handling unstructured flow data with irregular geometries. This system can be described by Eq.~\ref{eq:ns} with a stochastic forcing term $\bm{f} = [f_x(\bm{x},t), f_y(\bm{x},t)]^T$, which is governed by a stochastic diffusion equation,
\begin{equation}
    \frac{\partial \bm{f}}{\partial t} = \nu_f \nabla^2 \bm{f} + \bm{\delta},
    \label{eq:elbow:force}
\end{equation}
where $\bm{\delta} = [\delta_x, \delta_y]^T$ represents a stochastic source term, with each component sampled from a standard normal distribution $\delta_x,\delta_y \sim \mathcal{N}(0,1)$, and $\nu_f = 2$ is the diffusion coefficient for spreading the stochastic forcing. To generate training data, DNS is conducted by solving these stochastic incompressible NS equations on a 2D irregular domain with unstructured grids (see Fig.~\ref{fig:elbow}(d)). A long-span spatiotemporal flow sequence $\bm{\Phi}^{dns}$ consisting of $16,000$ instantaneous flow fields of $u, v, p$ is obtained from the DNS, subsequently partitioned into $15,873$ shorter sub-sequences $\tilde{\bm{\Phi}}_i$, each consisting of $N_t = 128$ snapshots, to assemble a dataset, of which 80$\%$ is used for training ($\mathcal{A}_{train} = \{\tilde{\bm{\Phi}}_i\}_{i=1}^{12,698}$) and remaining is reserved for testing purpose. 
% (for more details on data generation, refer to Sec.~\ref{sec:app:elbow})
 The CoNFiLD is trained on $\mathcal{A}_{train}$ unconditionally.    

The results generated by the CoNFiLD model are compared with DNS references in Fig.~\ref{fig:elbow}. Panel (a) depicts a sequence of velocity magnitude snapshots from DNS at the $0^{th}$, $320^{th}$, $640^{th}$, $960^{th}$, and $1280^{th}$ numerical time steps, showcasing the stochastic spatiotemporal dynamics through irregular vortex movement patterns over time. For comparison, three randomly generated flow sequence samples by CoNFiLD are presented in panel (b), which exhibit similar stochastic behaviors, maintaining visual and physical consistency with coherent temporal evolution and clearly defined boundary layers. Despite their similar stochastic spatiotemporal behavior, the instantaneous flow patterns differ across different generated trajectory samples and the DNS reference, highlighting CoNFiLD's ability to capture the underlying distribution of the training dataset instead of merely replicating label data. This is further substantiated in Fig.~\ref{fig:elbow}(c), through a comparison of the probability density function (PDF) of velocity magnitude and pressure between the 25 CoNFiLD generated flow sequences and the DNS datasets. The PDFs of velocity magnitude (left panel) and pressure (right panel) for both CoNFiLD-generated samples (red curves) and DNS data (blue curves) show close alignment, with only minor discrepancies observed at certain peaks of the velocity magnitude PDF. In Fig.~\ref{fig:elbow}(e), the time-averaged velocity magnitude, $\displaystyle M(x,y) = \frac{1}{N_t}\sum_{t=1}^{N_t}||\bm{u}_t(x,y)||$, derived from the CoNFiLD-generated samples is almost identical to that of the reference DNS data, with an average discrepancy value of merely $0.041$, representing approximately $4\%$ difference from the reference mean. Figure~\ref{fig:elbow}(f) presents the standard deviation of velocity magnitude over time, $S(x,y) = \sqrt{\frac{1}{N_t} \sum_{i=1}^{N_t} \left(||\bm{u}_t(x,y)|| - M(x,y)\right)^2}$, for generated samples against reference DNS data. The minimal discrepancy in standard deviation, with an absolute mean spatial discrepancy of $0.0294$—approximately $8.4\%$ of the reference, demonstrates CoNFiLD’s capability to not only generate accurate spatiotemporal samples but also effectively capture the underlying distributions.

\subsubsection{Generating equilibrium inflow turbulence of 3D channel flows} \label{sec:turb_inflow}
In this subsection, we demonstrate the CoNFiLD model on synthesizing sequences of instantaneous inlet velocity fields for 3D turbulent channel flows, highlighting its utility in generating accurate inflow turbulence boundary conditions, critical for eddy-resolving simulations. Focused on a fully-developed turbulent channel flow, governed by the incompressible NS equations with a forcing term $\bm{f}$ that simulates constant pressure gradients driving the flow, this setup ensures homogeneity in the streamwise and spanwise directions, while turbulence statistics exhibit variations only in the wall-normal direction~\cite{pope2000turbulent}. Our objective here is to generate time-coherent, three-dimensional instantaneous velocity fields at the channel's $z-y$ cross-section ($\bm{u}(y, z, t) = [u(y, z, t), v(y, z, t), w(y, z, t)]^T$: $\partial\Omega \times \mathbb{R}^+ \to \mathbb{R}^3$). The training data, obtained from fully-resolved DNS of a 3D turbulent channel flow at a friction Reynolds number of $Re_\tau = 180$, is sampled over a duration of four flow-through-time ($T_{\mathrm{flow}}$) with a learning step size of $\Delta t^+_{\mathrm{train}}=0.4$ that is $100\times$ numerical time step size $\delta t = 0.004$, exhibiting temporal correlation. Only instantaneous velocity flow fields on one cross section $\bm{\Phi}^{dns}$ of $1,200$ learning time steps are collected to create our dataset. The same DNS resolution is maintained, i.e., $N_z \times N_y = 100 \times 400$. The DNS flow sequence is divided into $945$ shorter sub-sequences $\tilde{\bm{\Phi}}_i$, with each comprising $N_t = 256$ snapshots roughly corresponding to one $T_{\mathrm{flow}}$. This forms a database, of which $80\%$ is used as the training set $\mathcal{A}_\mathrm{train} = \{\tilde{\bm{\Phi}}_i\}_{i=1}^{756}$ and the remaining $20\%$ is reserved as the test set $\mathcal{A}_\mathrm{test} = \{\tilde{\bm{\Phi}}_i\}_{i=1}^{189}$ in the conditional generation. 
%Further details on data generation are provided in Section \ref{sec:app:inlet}.

\begin{figure}[t!]
    \centering
    \includegraphics[width=0.9\textwidth]{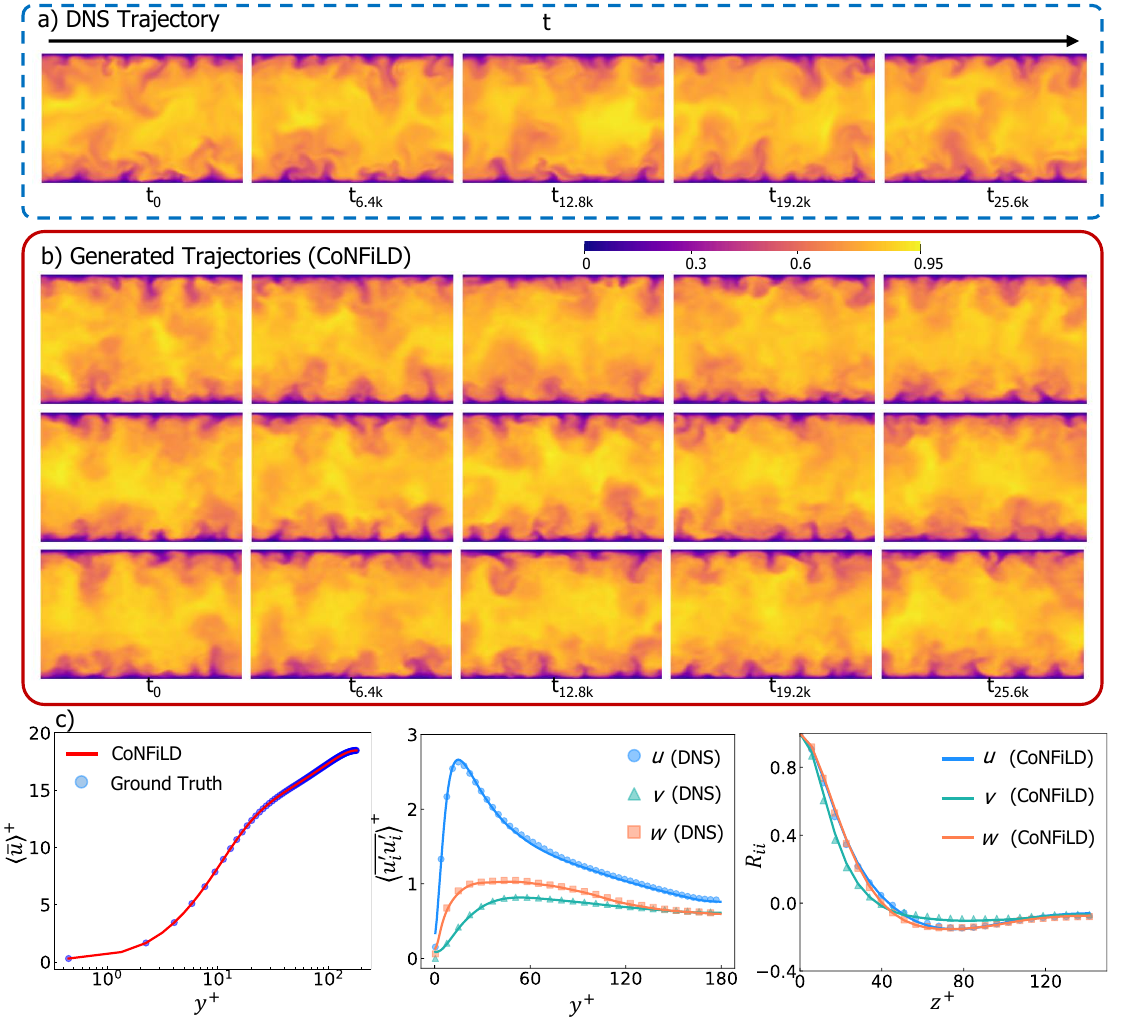}
    \caption{Unconditional generation of equilibrium inflow turbulence. (a) Instantaneous streamwise velocity $u$ obtained by DNS. (b) Three distinct realizations of $u$ generated by CoNFiLD. (c) Analysis of turbulence statistics, highlighting mean streamwise velocity (left), root-mean-square (RMS) of velocity fluctuations (middle), and two-point correlations of each velocity component at $y^+=150$ (right). $\overline{\Box}$ indicates time-averaged quantities, while $\langle \Box \rangle$ denotes ensemble average across all samples. Spatial coordinates are normalized by the wall unit $y^+=\frac{y u_{\tau}}{\nu}$, where $y$ is the wall normal distance, $u_\tau$ is the friction velocity, and $\nu$ is the kinematic viscosity. Velocity statistics are scaled by $u_\tau$ for normalization. }
    %(d) energy spectrum of velocity at $y^+=150$,The energy in (d) is normalized by wall unit. 
    \label{fig:inlet contour}
\end{figure}
The unconditional inflow turbulence generation results of CoNFiLD are compared with DNS reference in Fig.~\ref{fig:inlet contour}, illustrating both the fidelity and diversity of the CoNFiLD-generated spatiotemporal velocity field samples. For this assessment, an ensemble of 50 flow sequences, each with 256 snapshots (equivalent to 25,600 numerical steps), was synthesized to ensure statistical convergence. Out of these, three exemplary flow sequences generated by CoNFiLD are showcased in Fig.~\ref{fig:inlet contour}(b), where the stochastic behavior and vortex patterns all visually resemble those of the DNS reference in Fig.~\ref{fig:inlet contour}(a), affirming the model's fidelity in capturing the essence of turbulent flows. Notably, the individual instantiations of the generated flow fields exhibit substantial variability, showcasing a departure from the deterministic nature of neural solvers like ConvLSTM or Transformer architectures~\cite{han2022predicting}, which are conventionally engineered to output a single deterministic realization. This comparison underscores CoNFiLD's ability to not only capture the complex dynamics of turbulent flows but also to introduce a rich diversity in the synthesized spatiotemporal velocity field samples, a critical aspect for the realistic representation of turbulence phenomena. To further quantitatively evaluate the performance of the CoNFiLD model, we conducted a detailed analysis of the turbulence statistics across all generated flow sequence samples. As shown in Fig.~\ref{fig:inlet contour}(c), the turbulence statistics obtained from our model are in good agreement with those obtained by DNS. In particular, the mean streamwise velocity profile generated by CoNFiLD accurately matches with the DNS, reflecting the expected behavior across the linear viscous sublayer, buffer layer, and logarithmic law region. Similarly, the root-mean-square (RMS) of velocity fluctuations generated by CoNFiLD is in good agreement with the DNS results. Additionally, the two-point correlation exhibits an initial decline to negative values before asymptotically approaching zero, aligning with DNS observations. This analysis demonstrates that CoNFiLD-generated flow captures the entire range of turbulence scales and structures, resembling those identified in DNS with remarkable accuracy. Notably, we didn't find any discernible bumps and wiggling in the two point correlations of generated flow as reported in Gao et al.~\cite{gao2023bayesian}, showing CoNFiLD's superior performance compared to state-of-the-art generative methods such as the video diffusion model.

\subsubsection{Generating non-equilibrium turbulence of periodic hill}
\label{sec: uc-ph}

In addition to the previous scenario, we further demonstrate CoNFiLD's capability in generating spatiotemporal non-equilibrium turbulence flows through a classical periodic hill benchmark case, featuring a broad spectrum of complex flow behaviors including separation, recirculation, and reattachment. These complex turbulence phenomena are prevalent in a wide range of engineering applications, from aerospace propulsion to chemical processing, and pose significant challenges for both traditional numerical models and data-driven surrogates~\cite{breuer2009flow}. For the periodic hill case, the turbulence is statistically two-dimensional -- in streamwise ($x$-) and wall-normal ($y$-) directions. Therefore, the CoNFiLD here is trained to generate time-coherent, three-dimensional instantaneous velocity fields at the $x-y$ plane, $\bm{u}(x, y, t) = [u(x, y, t), v(x, y, t), w(x, y, t)]^T$: $\mathbb{R}^2 \times \mathbb{R}^+ \to \mathbb{R}^3$. Similar to the previous example, the training data is a subset of fully-resolved 3D DNS simulation results with $Re_h = 2800$, defined by the height $h$ of the hill. Specifically, to manage computational costs, we first downsample the 3D DNS data over a duration of $10 T_{\mathrm{flow}}$ using a learning time step size of $\Delta t^+_{\mathrm{train}}=1.9$, which consists of $300$ numerical timesteps, retaining the temporal coherence. We then select three spanwise cross sections along the $z$ axis from the downsampled data, spaced apart by a distance of $\Delta z^{+}_{\mathrm{slice}}=112$, thereby reducing spatial correlation. Fourier Fast Transform (FFT) filter is applied to eliminate high frequencies beyond a certain threshold and reduce the spatial resolution to $N_x \times N_y = 88 \times 133$, with the threshold set by the highest frequency the downsampled mesh can accurately represent, according to the Nyquist-Shannon sampling theorem. 
\begin{figure}[t!]
    \centering
    \includegraphics[width=1.0\textwidth]{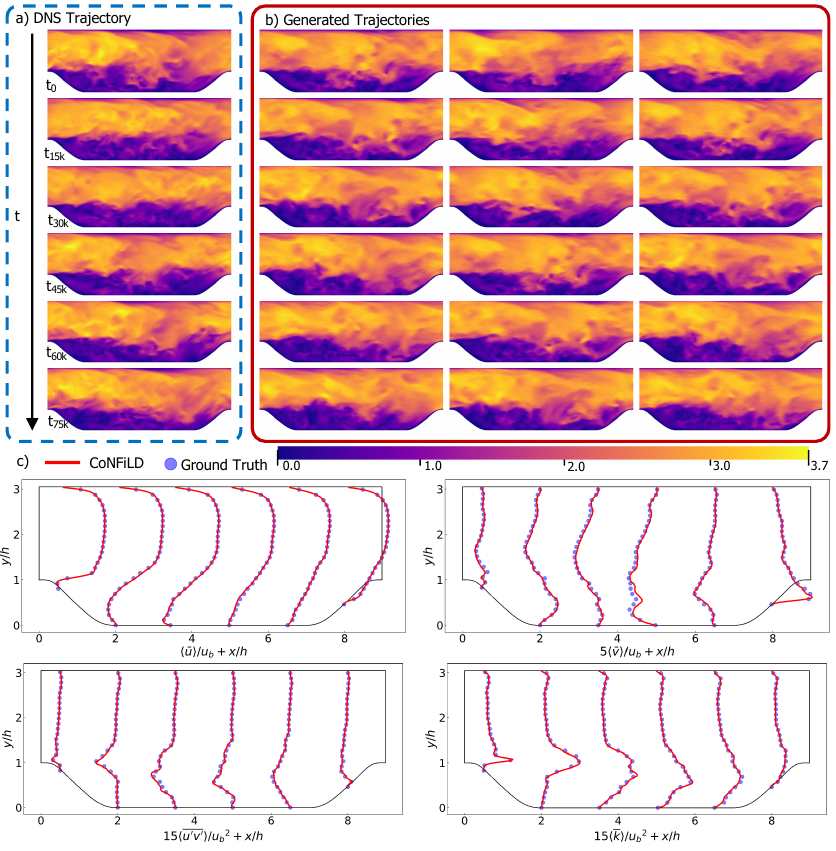}
    \caption{Unconditional generation of non-equilibrium turbulence over periodic hills. (a) Instantaneous velocity magnitude of the DNS flow data. (b) Instantaneous velocity magnitudes of three randomly generated realizations by CoNFiLD. (c) Turbulence statistics at selected locations, including the mean streamwise velocity $\bar{u}$ (upper left), mean vertical velocity $\bar{v}$  (upper right), Reynolds shear stress $\overline{u'v'}$ (lower left), and the total turbulence kinetic energy ($\bar{k}$) (lower right). The spatial coordinates are normalized by the hill height $h$, and the statistical quantities are normalized by the bulk velocity $U_b$.  }
    \label{fig:PH contour}
\end{figure}
This extensive flow sequence $\bm{\Phi}^{dns}$ is partitioned into $2,115$ shorter sub-sequences $\tilde{\bm{\Phi}}_i$, each containing $N_t = 256$ snapshots. This forms a dataset $\{\tilde{\bm{\Phi}}_i\}_{i=1}^{2115}$, where $80\%$ is used for training $\mathcal{A}_\mathrm{train} = \{\tilde{\bm{\Phi}}_i\}_{i=1}^{1692}$ and $20\%$ is reserved for testing $\mathcal{A}_\mathrm{test} = \{\tilde{\bm{\Phi}}_i\}_{i=1}^{423}$.
%Further details regarding the generation of label data and the configuration of periodic hills are provided in~\ref{sec:app:ph}.

The comparison of instantaneous flows unconditionally generated by CoNFiLD against the ground truth, derived from DNS data, is shown in Fig.~\ref{fig:PH contour}, where velocity contours and turbulence statistics are analyzed. Three of the 150 CoNFiLD-generated spatiotemporal trajectories are randomly selected and presented in panel (b), each comprising a total of $N_T = 1024$ snapshots (equivalent to $307,200$ numerical steps), against the DNS ground truth in panel (a). The comparison shows that all the CoNFiLD-generated flow samples vividly recreate similar vortex structures and flow characteristics of this non-equilibrium turbulent flow as the reference, showcasing CoNFiLD's exceptional ability to synthesize realistic and physically accurate non-equilibrium turbulent behaviors. Similar to the prior example, each generated sample retains uniqueness while closely mimicking the physical behavior of ground truth. The physical validity of generated flows is further quantitatively evidenced by the statistical analysis presented in Fig.~\ref{fig:PH contour}(c), where the time-averaged mean flow profiles of the generated flow sequences align closely with the labeled data in both streamwise and wall-normal directions. Detailed examination of the velocity profiles identifies a consistent pattern of flow separation immediately downstream of the hill (at $x/h \leq 5$) across all generated samples, mirroring the DNS results. Additionally, both the generated and DNS data exhibit a clear recirculation zone between $x/h=2$ and $x/h=4$ with reattachment occurring around $x/h = 5\sim5.5$, where no negative mean velocity is observed. More remarkably, the Reynolds shear stress $\langle \overline{u'v'}\rangle$ and turbulence kinetic energy (TKE) $\bar{k}$ (=$\displaystyle\frac{1}{2}\left(\langle \overline{u'u'}\rangle +\langle \overline{v'v'}\rangle \right)$ of the generated samples closely match those of the ground truth, with peaks observed in the free shear layer (shown in Fig.~\ref{fig:PH contour} (c)). These results affirm that the turbulence synthesized by CoNFiLD faithfully replicates the statistical characteristics of the label data. Notably, conventional RANS and LES methods tend to underpredict some of these statistical metrics, especially in complex flow regimes with separations and recirculations. In contrast, CoNFiLD can accurately capture the flow statistics yet with substantially less computational cost, as further discussed in Sec.~\ref{sec:discussion}.

\subsubsection{Generating 3D wall-bounded turbulence with wall roughness}
\label{sec:roughwall}

After showcasing CoNFiLD's effectiveness in synthesizing cross-sectional spatiotemporal turbulence, we extend its application to a more challenging scenario: the spatiotemporal generation of sophisticated instantaneous wall-bounded turbulent flows within 3D domains featuring regular wall-roughness elements. Turbulent flows over a rough surface are ubiquitous in various naval systems due to manufacturing processes or service-induced erosion and biofouling \cite{KADIVAR2021100077}. Different roughness conditions significantly affect near-wall turbulence structures and the transfer of scalar, momentum, and energy, impacting the safety, performance, and efficiency of marine systems. However, accurately modeling and predicting rough-wall turbulence with eddy-resolving simulations demand prohibitive computational resources, positioning CoNFiLD as a valuable alternative for fast surrogate modeling. In response, CoNFiLD is applied in this case to learn from high-fidelity DNS data, enabling the efficient generation of realistic turbulent flows over rough surfaces with significant speedup.
%present a higher-level capability---to generate practical and sophisticated instantaneous 3D turbulence, demonstrated by a wall-bounded turbulence case. Such authentic simulations of turbulent flow are crucial across various natural and engineered domains, from atmospheric and ocean currents to the internal dynamics of devices like wind turbines, artificial hearts, and propulsion systems. Conventional computational methods and machine-learning surrogates often falter with these complex phenomena, positioning CoNFiLD as a valuable alternative for these challenges. 
Specifically, our goal here is to generate time-coherent, four-dimensional realistic instantaneous velocity fields ($\bm{u}(x, y, z, t) = [u(x, y, z, t), v(x, y, z, t), w(x, y, z, t)]^T$: $\Omega \times \mathbb{R}^+ \to \mathbb{R}^3$). 
The training data originates from a fully resolved 3D transient DNS of wall-bounded turbulence over cubic roughness elements, at a Reynolds number of $Re_h=3200$, which is based on the cube height $h$.
%{, as detailed in Section~\ref{sec:app:roughwall}.} 
We subsample exclusively during the fully developed phase of flow using a time step of $\Delta t^+_{\mathrm{train}}=0.8$, which is $100 \times$ the numerical timestep, to preserve temporal correlation. Due to the GPU memory limitations in our lab, the training and turbulence generation for this 3D domain focus to a sub-region. 
%as marked in Fig.~\ref{fig:cube-config}. 
We apply the same filtering and downsampling methods as detailed in Sec.~\ref{sec: uc-ph} for the 3D sub-region, resulting in a spatiotemporal flow sequences $\bm{\Phi}^{dns}$ consisting of $1200$ snapshots with a resolution of $N_x \times N_y \times N_z=32 \times 34 \times 62$. This long-span sequence is partitioned into $817$ shorter sub-sequences $\tilde{\bm{\Phi}}_i$, each consisting of $N_t = 384$ snapshots, to assemble the dataset $\{\tilde{\bm{\Phi}}_i\}_{i=1}^{817}$. During training, $80\%$ of the database is used as the training set $\mathcal{A}_\mathrm{train} = \{\tilde{\bm{\Phi}}_i\}_{i=1}^{653}$ and the remaining $20\%$ is reserved as the test set $\mathcal{A}_\mathrm{test} = \{\tilde{\bm{\Phi}}_i\}_{i=1}^{164}$ for conditional generation validation. 

\begin{figure}[hp!]
    % \centering
    \includegraphics[width=1.0\textwidth]{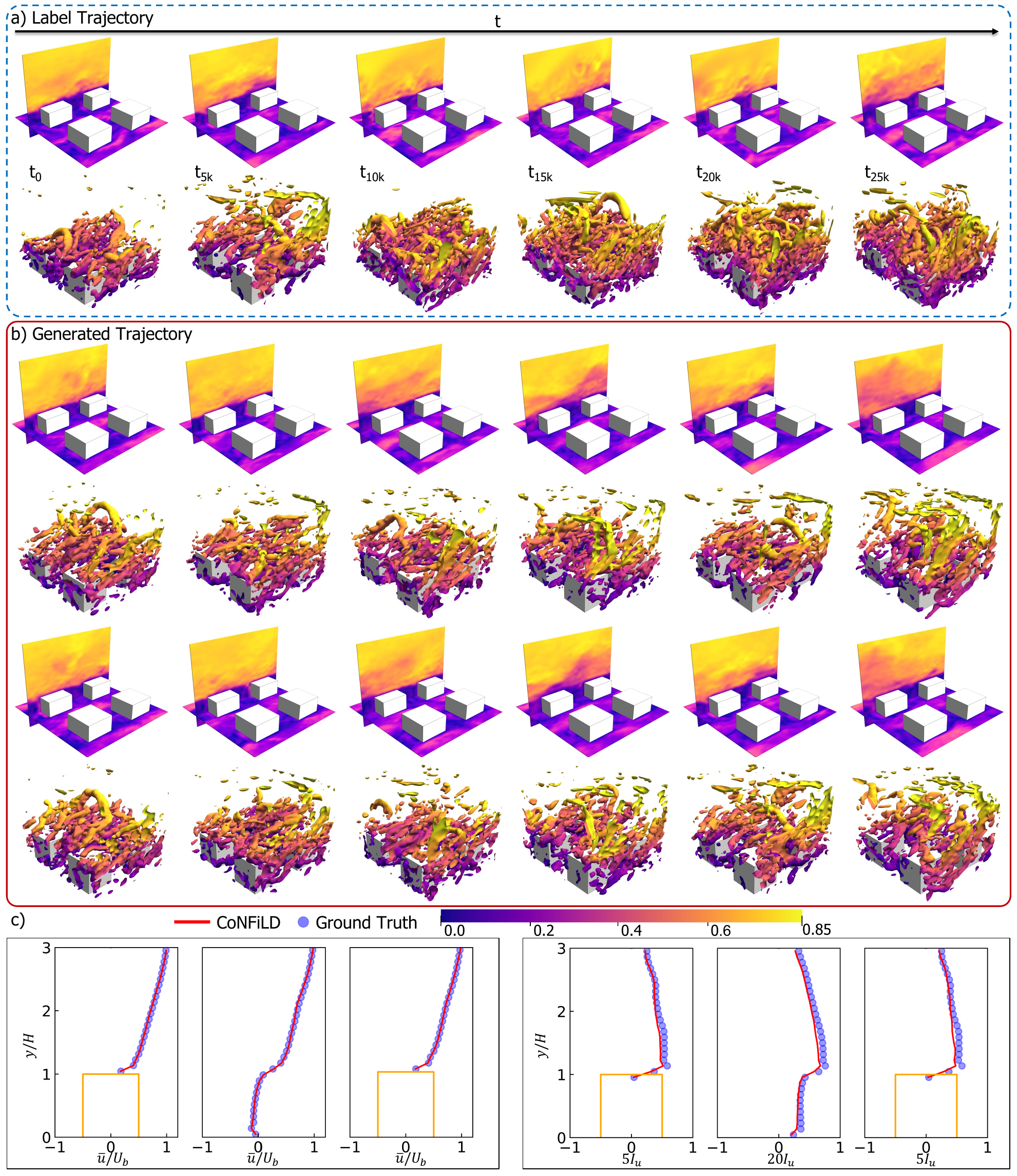}
    \caption{Unconditional generation of 3D wall-bounded turbulence. The instantaneous velocity magnitude (top) and iso-surfaces of Q-criterion (bottom)  of (a)  DNS and (b) two randomly generated realizations by CoNFiLD. (c) The turbulence statistics, including the time-averaged streamwise velocity $\bar{u}$ (left), and turbulence intensity $I_u$ (right), both above and between the roughness elements. Spatial coordinates are normalized by the height of the roughness elements, and the statistics are normalized by bulk velocity $U_b$.}
    \label{fig:3D contour}
\end{figure}
Figure~\ref{fig:3D contour} (b) showcases two instances of flow generated unconditionally by CoNFiLD, encompassing 1536 learning steps, equivalent to 153,600 numerical timesteps, alongside the labeled flow trajectory depicted in Fig.~\ref{fig:3D contour} (a). These flow trajectory samples are visualized through the velocity magnitude contours and the isosurfaces of the Q criterion, providing a detailed view of the three-dimensional turbulence characteristics within the domain. Notably, our CoNFiLD accurately reproduces the large-scale vortices associated with the roughness, closely mirroring the dynamics observed in DNS. Meanwhile, noticeable differences in the small-scale vortices among the generated samples and DNS highlight CoNFiLD's capability to capture the inherent probabilistic nature of wall-bounded turbulence. Additionally, the first- and second-order turbulence statistics of flows generated by CoNFiLD and those obtained from DNS are compared in Fig.~\ref{fig:3D contour} (c), featuring both time-averaged velocity and turbulence intensity at three representative locations. The agreement of flow statistics between CoNFiLD and DNS demonstrates the model's efficacy in vividly reproducing the instantaneous unsteady flow patterns, which preserve the accurate mean flow characteristics, indicating a successful replication of the primary flow mechanism. The results underscore the model's proficiency in generating varied instances of wall-bounded turbulence over extended duration beyond the training scope, providing statistical and physical fidelity superior to traditional RANS or unsteady RANS, which often fails to accurately predict flow separations and reattachments around roughness elements.~\cite{santiago2007cfd}.

\subsection{Zero-shot conditional spatiotemporal generation without retraining}
In addition to generating diverse flow realizations that adhere to the underlying distribution learned during its training phase, the trained CoNFiLD model is also capable of producing specific flow realizations conditioned on given inputs, without the need for retraining. This feature significantly highlights our model's versatility, enabling efficient and tailored flow predictions for various application scenarios. In this subsection, three different conditional generation applications -- sensor-based flow reconstruction, flow data restoration, and super-resolved generation -- are showcased and discussed. 

\subsubsection{Flow reconstruction from sparse sensor measurements}
We first explore an application of significant practical importance: full-field spatiotemporal reconstruction of flow from sparse sensor data through zero-shot conditional generation, underpinned by Bayesian posterior sampling. This capability is essential across various engineering domains, where obtaining comprehensive full-field flow information is challenging due to complex setups, prohibitive computational costs, or the inherent sparsity and noise in direct measurements. Traditional approaches have primarily adopted deterministic models, incorporating dimensionality reduction techniques like POD or DNN-based autoencoders~\cite{dubois2022machine,callaham2019robust,erichson2020shallow}. Although these methods have demonstrated some success in flow reconstruction, they often struggle with accuracy, robustness, and scalability, particularly in large-scale, complex turbulent flow scenarios~\cite{sofos2022current}.

\begin{figure}[tp!]
    \centering
    \includegraphics[width=1.0\textwidth]{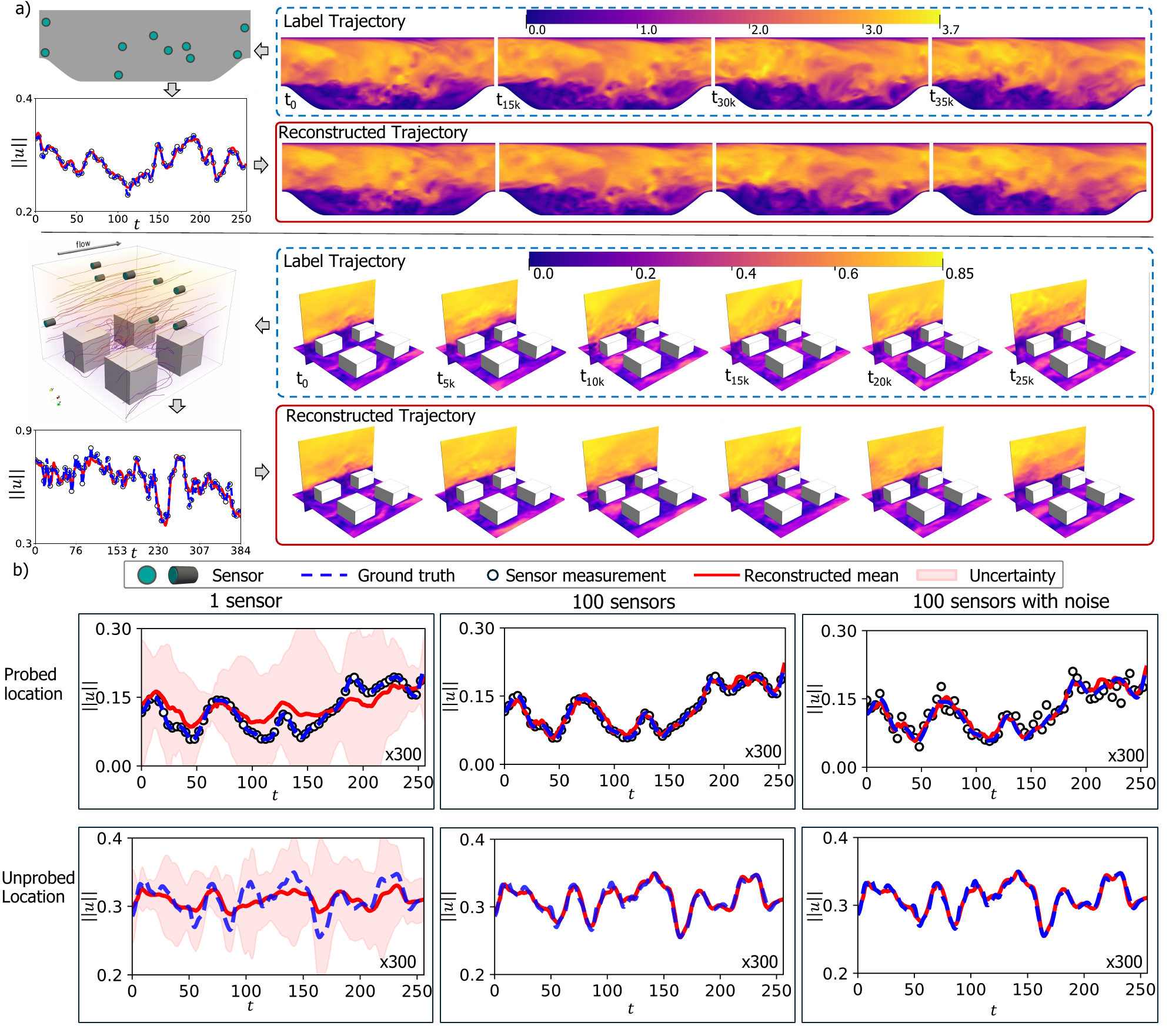}
    \caption{Flow reconstruction from limited sensor measurements using zero-shot conditional generation. (a) Comparison of label trajectory and the reconstructed on for non-equilibrium turbulence over periodic hill (1st and 2nd rows) and 3D wall-bounded turbulence (3rd and 4th rows), where right panel shows the contours and left panel shows sensor locations and single-point time-series signals at one sensor location. (b) Sensitivity study for number of sensors (1st column v.s. 2nd column) and w/o noise (2nd column v.s. 3rd column) at probed points (first row) and unprobed location (2nd row) for periodic hill case.}
    \label{fig:sensor}
\end{figure}
We demonstrate CoNFiLD's sensor-based conditional generation capability on the two non-equilibrium wall-bounded turbulence cases: flow over periodic hills and wall roughness elements, as presented in Secs~\ref{sec: uc-ph} and~\ref{sec:roughwall}. The problem is formulated as following: placing limited number of flow sensors sparsely within the flow field simulated by DNS to collect velocity signals at different times ($\bm{\Psi}$). These measurements serve as conditional inputs for CoNFiLD to generate full-scale spatiotemporal fields of this specific flow realization that is observed. 
For the periodic hill case, we randomly selected a flow sequence $\bm{\Phi}$ from the test dataset ($\bm{\Phi} \in \mathcal{A}_{\mathrm{test}}$) as the ground truth, containing $N_T=256$ snapshots, equivalent to $76,800$ numerical steps. Similarly, for the 3D wall-roughness case, the ground truth is a randomly selected test flow sequence of $N_T=384$, corresponding to 38,400 numerical steps. For the periodic hill and wall roughness cases, we randomly placed 10 and 100 sensors, corresponding to $0.1\%$ and $0.17\%$ of the grid points in each case, respectively. These sparse sensor measurements are then utilized to reconstruct the full-field spatiotemporal flows. Performance is assessed by comparing the reconstructed flows to the ground truth, as shown in Figure~\ref{fig:sensor}(a), which displays contour comparisons and single-point time-series signal analysis at the sensor location. Unlike unconditional generation, the reconstructed flows, despite being one of many realizations generated by CoNFiLD, show notable similarities to the ground truth in both contour maps and sensor signal patterns, owing to the inclusion of conditional information (i.e., sensor measurements). Note that the conditionally generated samples, though very similar, are still slightly different from each other. The scattering of the generated ensemble can be viewed as the uncertainty of the flow reconstruction, bypassing the necessity for model retraining. This adaptability and stochasticity of CoNFiLD enable it to not only reconstruct the specific flow realization observed by the sensors but also provide uncertainty estimates accordingly. This capability distinctively differentiates our approach from deterministic regression-based reconstruction methods, which are restricted to producing a single deterministic flow sequence. A closer examination of the contours indicates minor discrepancies in capturing small-scale flow structures, consistent with unconditional generation. The disparity is slightly more noticeable in the 3D rough-wall turbulence case, reflecting its higher complexity. Future improvements in model capacity and computational resources may address these limitations.     

We further explored how sensor configuration influences flow reconstruction performance in the periodic hill case. This involved adjusting the number of sensors and incorporating noise to better simulate real-world conditions, with the results presented in Fig.~\ref{fig:sensor}(b). The first and second columns compare the reconstruction performance using 1 and 100 sensors, respectively, by plotting mean and standard deviation (std) together with the ground truth for both probed and unprobed locations.  To ensure statistical reliability, we generated and analyzed 50 samples, determining the mean and std. The uncertainty is visualized by shading an area that spans three stds from the mean, providing clear insight into the variability of generated realizations. With a single sensor, the reconstructed uncertainty is considerable; however, the mean, despite deviating from the ground truth, roughly follows its trend at both probed and unprobed locations. Increasing the sensor count to 100 significantly enhances the alignment of the mean curve with the actual data and markedly narrows the uncertainty bounds. This improvement aligns with the expectations from Bayesian perspective, as more conditional information sharpens the high-density regions of the likelihood function, resulting in a more concentrated posterior distribution. Intuitively, our certainty about the reconstructed flow field increases with more observations. Additionally, we introduced Gaussian noise ($10\%$ of the original data range) to the signal of the 100 sensors and plotted the results in the third column. Compared to the second column, there is no notable performance drop at both probed and unprobed points even with noisy measurements, indicating the robustness of our model. These findings underscore the significant potential of our CoNFiLD model in scaling up to various real-world applications, demonstrating its flexibility with respect to sensor arrangements and its robustness against variations in signal quality.

\subsubsection{Flow restoration from damaged data}

The storage of turbulence data presents a substantial challenge within the CFD community, with data corruption noted as a major concern~\cite{brown2010scientific}. Although physics-based~\cite{Lee_2015, han2022reconstruct} and deep learning strategies~\cite{de2019data} have shown success in recovering fluid dynamics data for canonical flows, such as lid-driven cavity and flow around a cylinder, their applications in restoring turbulent flow data is less explored. To tackle this problem, we demonstrate another novel application of CoNFiLD:  high-fidelity restoration of corrupted turbulence data. We use the damaged data as conditional input ($\bm{\Psi}$) to facilitate the recovery of lost flow information by conditional generation. In this study, the data damage is defined as the absence of flow information at a central subregion of the fluid domain, mathematically described as a spatiotemporal masking operation. This objective is to precisely restore the missing flow details by leveraging the information available from the surrounding regions.

\begin{figure}[tp!]
    \centering
    \includegraphics[width=\textwidth]{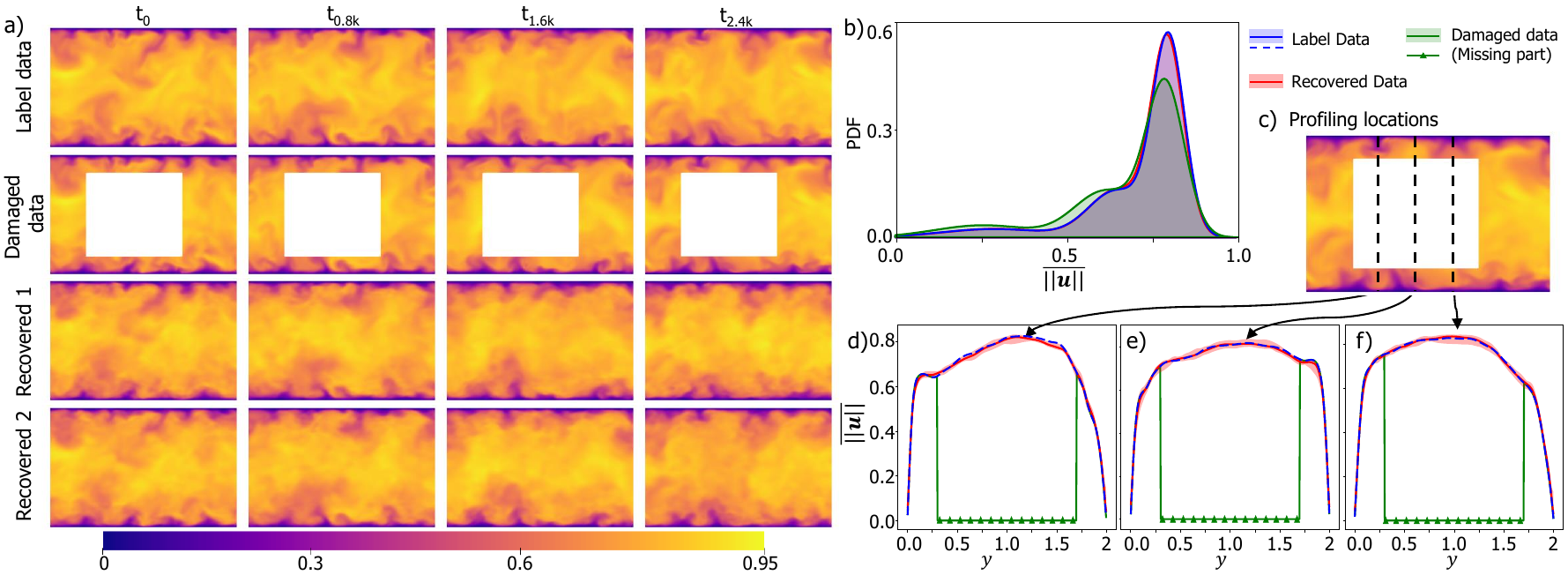}
    \caption{Inpainting of inlet turbulence for channel flow. (a) Instantaneous velocity magnitude contours of DNS, damaged and recovered data. (b) Comparison of the PDF of the velocity magnitude between DNS, damaged and recovered data. (c) Comparison of the velocity magnitude profile between DNS, damaged, and recovered data at three spanwise locations}
    \label{fig:inpainting}
\end{figure}
% Case $2$
Using the turbulence inlet case previously presented in Section~\ref{sec:turb_inflow}, we illustrate the data restoration capability of the CoNFiLD model. A subset of the trajectory (${\bm{\Phi}}$) with $N_T=32$ frames (equivalent to $3,200$ numerical time steps) from the test dataset, previously unseen by the CoNFiLD model, is selected as the ground truth. The corrupted data are created by masking the central subregion of the ground truth across all time steps, as shown in the second row of Fig.~\ref{fig:inpainting}(a). These corrupted data then serve as conditional information for CoNFiLD to infer the flow dynamics within the masked area. Notably, the square damaged region defined here is illustrative; in practice, the shape of the damaged region can vary significantly, extending to the domain's boundaries without restrictions.

To accurately quantify uncertainty in the restoration process, we generate $18$ conditioned samples, two of which are presented in Fig.~\ref{fig:inpainting}(a) alongside the original and damaged data. The CoNFiLD model consistently restores the flow within the masked areas, seamlessly integrating with the surrounding data without noticeable discrepancies/inconsistencies at the interface. However, each generated sample varies slightly from the others, subtle in the contour plots but apparently reflected in the depiction of uncertainty regions shown in Fig.~\ref{fig:inpainting}(d). A closer look at velocity magnitude $||u||$ profiles at three cross-sections (Fig.~\ref{fig:inpainting}(d)), reveals increased uncertainty from the periphery towards the center of the damaged area. This trend is due to higher spatial covariance with adjacent known flow information near the edges, leading to reduced uncertainty compared to the central portion of the masked area. Nonetheless, the overall uncertainty remains minimal, suggesting that CoNFiLD effectively utilizes surrounding flow information to draw from the posterior distribution closely aligned with the ground truth. Further evidence of CoNFiLD's proficiency is presented in Fig.~\ref{fig:inpainting}(c), where it significantly refines the probability density function (PDF) of the velocity magnitude $||u||$ of the damaged data, aligning it closely with the PDF of the original data.

\subsubsection{Spatiotemporal super-resolution of low-fidelity data}

Super-resolution techniques are rapidly being adopted across various computational and experimental communities to derive significant details from low-resolution (LR) images and data. Analytical, physics-based, and deep learning super-resolution techniques have shown promising results, from improving low-fidelity simulation results to enhancing under-resolved 4D flow MR imaging data ~\cite{geneva2020multi, gao2021super, shu2023physics, de2016temporal, toger2020blood, ferdian20204dflownet, rutkowski2021enhancement}. Motivated by these advancements, we present another capability of our proposed CoNFiLD model---creating highly detailed instantaneous flows from LR counterparts, showcasing significant potential for large-scale super-resolution challenges. Through the turbulence channel flow case,  we demonstrate the zero-shot super-resolution capability of the trained CoNFiLD model, regardless of the quality of LR data. We select a sub-trajectory (${\bm{\Phi}}$) comprising $N_T=256$ frames (equivalent to $25600$ numerical time steps) from the test dataset to serve as the ground truth. Three different levels of LR data are generated by downsampling the high-resolution (HR) DNS ($400\times100$) to three different resolutions, $64\times16$, $16\times4$, and $4\times1$, to cover a spectrum of LR scenarios typically encountered in practice. 25 samples are generated for each LR scenario to ensure accurate estimation of the statistical metrics.

\begin{figure}[t!]
    \centering
    \includegraphics[width=1.0\textwidth]{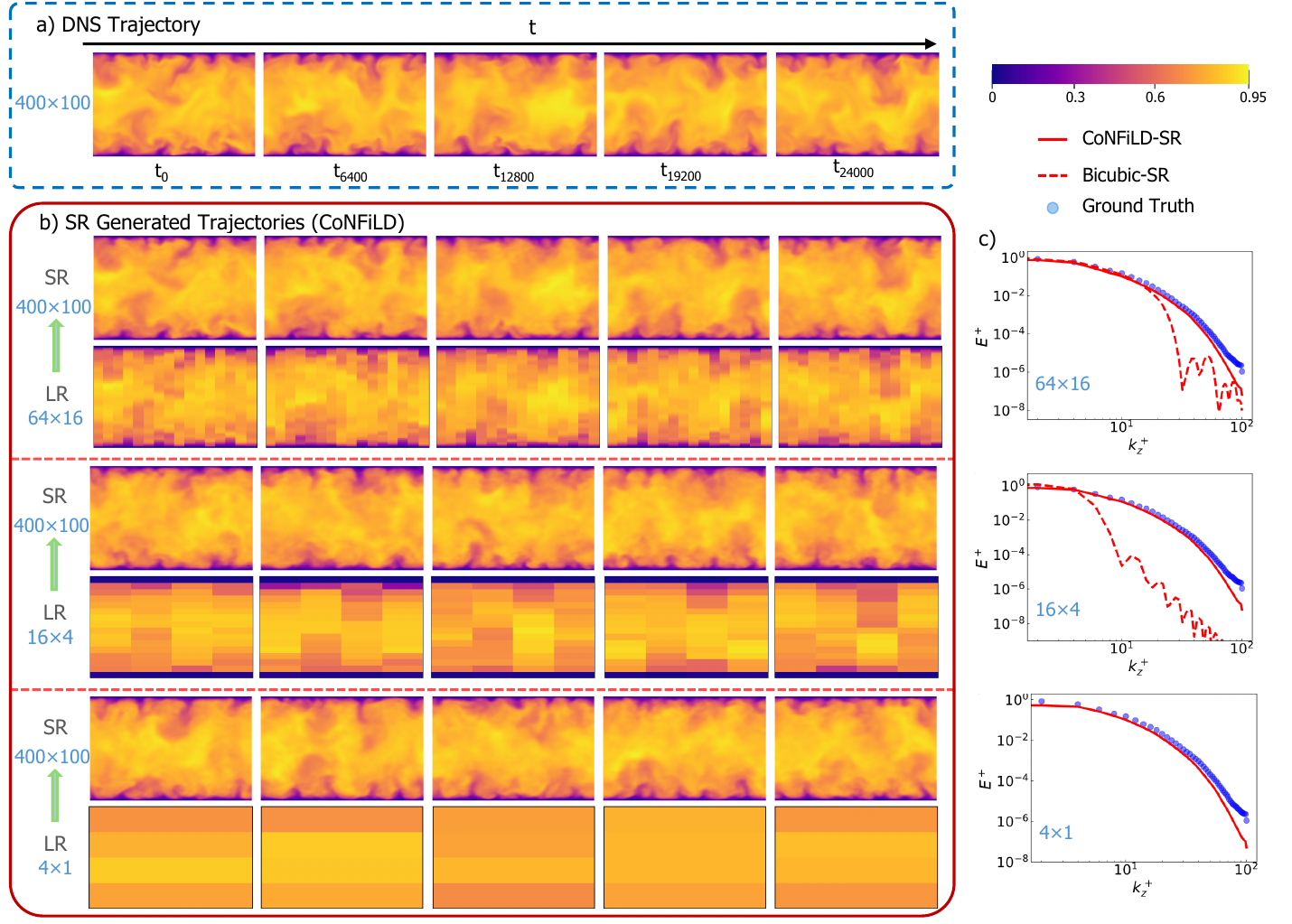}
    \caption{Super-resolution for inlet turbulence of 3D channel flow. (a) Instantaneous velocity magnitude contours of DNS. (b) Super-resolved (SR) generation results by CoNFiLD, comparing the conditional LR input and SR contours. (c) Comparison of TKE spectra cross DNS, CoNFiLD SR results, and bicubic interpolation baselines. Notably, due to its inadequate performance, bicubic interpolation is excluded for the most extreme downscaling scenario ($4 \times 1 \to 400\times100$).}
    \label{fig:inlet-sr}
\end{figure}
The performance of the CoNFiLD is illustrated in Fig.~\ref{fig:inlet-sr}, with panel (a) showing the ground truth trajectory ($400\times100$) and (b) displaying pairs of LR input and its super-resolved (SR) flow contours, across three different LR settings. Impressively, regardless of the initial quality of the LR data, all CoNFiLD-reconstructed flows are up-scaled to the original high resolution of $400\times100$, achieving a visual fidelity closely akin to the DNS reference. This can be further substantiated through the TKE spectrum analysis in Fig.~\ref{fig:inlet-sr}(c), comparing the SR flows against both the ground truth and the baseline SR result using bicubic interpolation. The bicubic SR method significantly fails to recover high-frequency details starting from such low resolution input, and its performance deteriorates with decreasing input quality. In stark contrast, CoNFiLD's reconstructions accurately replicate the true spectrum across all scales, even for the input with the lowest resolution ($4\times1$). Upon closer examination of the instantaneous flow contour comparisons, the SR reconstructions for the lowest resolution (at the bottom of Panel (b)) deviate from ground truth data, primarily because the exceedingly low input resolution provides negligible informative conditions, rendering the model's behavior similar to unconditional generation. As input resolution increases (from bottom to top in panel (b)), the conditionally generated SR samples increasingly align with the instantaneous flow patterns of the ground truth, with samples generated from the $64\times16$ LR input nearly indistinguishable from the DNS data. CoNFiLD's Bayesian formulation enables robust SR across varying input qualities, contrasting with many existing SR methods highly dependent on input resolution. Notably, CoNFiLD's ability to handle super-resolution tasks across different discretized flow representations—structured or unstructured—without retraining highlights its versatility. This adaptability clearly surpasses the capabilities of CNNs, which require retraining for different input resolutions and qualities. Similarly, although GNNs can manage unstructured data, they struggle with scale generalization. These attributes emphasize CoNFiLD's efficacy and adaptability in super-resolution applications, underlining its potential to tackle complex engineering challenges beyond flow data enhancement.

\section{Discussion}
\label{sec:discussion}
\begin{figure}[pt!]
    \centering
    \includegraphics[width=1.0\textwidth]{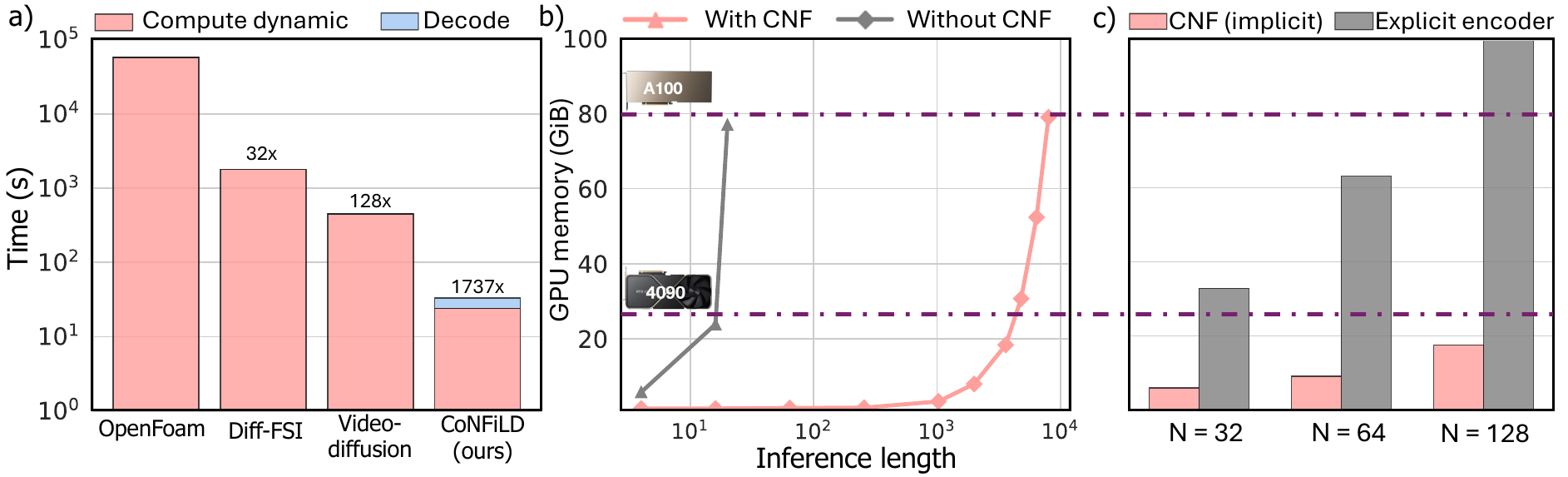}
    \caption{Performance evaluation. (a) Inference time cost comparison among OpenFOAM ran on CPU and Diff-FLowFSI ran on GPU, Video-diffusion model~\cite{gao2023bayesian}, and CoNFiLD (ours). (b) GPU memory cost comparison for unconditional generation w/o CNF encoder. (c) GPU memory cost comparison for implicit (ours) and explicit decoding strategy (e.g., POD, CNN) in subsampling-based conditional generation.}
    \label{fig:performance}
\end{figure}

To evaluate CoNFiLD's efficiency improvements over traditional CFD methods and existing DL-based generative models, we assessed its computational cost compared to established benchmarks, such as OpenFOAM (a CPU-based CFD solver in C++)~\cite{jasak2007openfoam}, Diff-FlowFSI (an in-house GPU-enabled, fully-vectorized differentiable CFD solver in JAX)~\cite{xiantao2024diffsi}, and the video diffusion model for spatiotemporal turbulence generation~\cite{gao2023bayesian}. This comparison was made using the inlet turbulence generation case detailed in Section~\ref{sec:turb_inflow}, reporting the expected time cost for generating $N_T=300$ (60,000 numerical timesteps) of spatiotemporal turbulence flow sequences in Fig.~\ref{fig:performance}(a). Compared to OpenFOAM, the GPU-accelerated fully vectorized JAX solver, Diff-FlowFSI achieves a 30-fold increase in speed. The video diffusion model, operating directly on physical space, further boosts this speedup to 128 times. Remarkably, by operating a diffusion process in latent space, our CoNFiLD extends this speedup to an impressive 1737-fold. This exceptional efficiency stems from multiple factors: Firstly, CoNFiLD runs on GPUs, contrasting with CPU-based OpenFOAM, providing an initial efficiency boost at the cost of higher memory demand. Secondly, compared to GPU-accelerated solvers like Diff-FlowFSI facing timestep constraints by the CFL condition, CoNFiLD can employ significantly greater timestep size without convergence issues. It leverages pre-trained knowledge on the probability distribution of all possible flow solutions for rapid online inference, offering an additional layer of efficiency boost. Thirdly, existing DL-based video generative modeling techniques such as the video diffusion model, though a similar probabilistic view, directly operate in pixel/physical space, which can be easily bottlenecked by memory for extended sequences. 
Unlike the model by Gao et al.~\cite{gao2023bayesian}, which requires an auto-regressive conditional generation for long-span generation, CoNFiLD generates latent images for large $N_T$ with a much smaller memory footprint, enabling direct generation without the need for auto-regressive sequential conditioning. The only overhead of the CoNFiLD model is the cost of the decoding process, which is negligible compared to the latent diffusion process (approximately ten times less). This distinction adds another layer of performance boost to our model. In summary, CoNFiLD achieves unparalleled performance gains among peer methods, showing substantial potential for scaling up to higher dimension flow data.

We further explored the memory usage differences when CoNFiLD performs diffusion processes either in physical space (without the CNF encoder) or in latent space (with the CNF encoder), focusing on unconditional generation scenarios for a fair comparison (shown in Fig.~\ref{fig:performance}(b)). This comparison also relates to other generative AI-based spatiotemporal flow generators that lack an encoding mechanism (e.g., video-diffusion~\cite{gao2023bayesian}). Monitoring CoNFiLD's memory demand over a range of inference lengths, from 1 to 8000 learning steps, both with and without the CNF encoder, we found significant performance differences. As illustrated in Fig.~\ref{fig:performance}(b), absent the CNF encoder, memory usage quickly reaches the maximum capacity of current top-tier GPUs with increasing inference length, maxing out at 16 learning time steps for the Nvidia RTX4090 and 20 steps for the Nvidia A100. In contrast, with the CNF encoder, memory consumption increases more gradually, enabling significantly longer inference stretches—up to 3900 and 8000 steps for the Nvidia RTX4090 and A100, respectively. This results in an extraordinary extension of inference lengths by factors of 243 and 400 for the two GPUs, underscoring the substantial benefits of latent space synthesis facilitated by the CNF encoder's robust encoding capabilities. Notably, the CNF achieves compression ratios of $0.017\%$ for periodic hill case and $0.002\%$ for 3D rough wall turbulence case, an impressive achievement given the complexity of the flows processed. While convolutional autoencoders (CAE) may reach similar compression ratios, as suggested by related research~\cite{pan2023neural,nakamura2021convolutional,hasegawa2020cnn}, the memory constraints of loading the full-field 3D/4D data become the bottleneck. Moreover, CNF's inherent implicit nature to handle unstructured data sets it apart---Unlike CAE, which explicitly encode data via convolutions and pooling on fixed regular grids, CNF allows CoNFiLD to train on and generate turbulence on unstructured grids simply by querying the CNF with desired coordinates and latent vectors.

% --------- JX Here -----------
Surprisingly, the advantages of incorporating CNF extend beyond this, as we found a distinct benefit for memory efficiency brought by the CNF in the subsampling-based conditional generation of CoNFiLD. In particular, the conditional generation process entails a forward evaluation and backpropagation of Eq.\ref{eq:latenttoobservation}. Note that this requires resolving the whole field before performing the forward function $\mathcal{F}$. The procedure remains the same if the CNF encoder is substituted by an explicit ML-based encoder like CAE. However, if the function $\mathcal{F}$ involves a subsampling process $\mathcal{M}$ in time and the spatial dimension, we can apply the subsampling $\mathcal{M}$ on the query spatiotemporal coordinates before passing them into the CNF decoder, thereby bypassing the recovery of the whole flow field and significant reducing the memory usage both in forward computation and backward gradient estimation. To demonstrate this, we define the forward function $\mathcal{F}$ simply as a masking function that preserves $10\%$ spatial points. The memory consumption of CoNFiLD using this pre-subsampling technique (only available with CNF) versus the original process (the only option for explicit encoders) for different inference lengths $N$ are plotted in Fig.~\ref{fig:performance}(c), where one can observe that using an explicit encoder quickly exceeds the top tier GPUs at $N=64$ and $N=128$ for Nvidia RTX4090 and A100 respectively. In contrast, the presence of CNF controls the memory cost under the limit of Nvidia RTX4090 in all three occasions. This fully verifies the memory benefit of CNF during the subsampling-based conditional generation of the CoNFiLD.

\section{Conclusion}
\label{sec:conclusion}
In this study, we introduce CoNFiLD, an innovative deep generative learning framework designed for probabilistic generation of complex, three-dimensional spatiotemporal turbulence. At its core, CoNFiLD uniquely combines a Conditioned Neural Field (CNF) with a latent diffusion model to enable scalable, long-span spatiotemporal generation. Specifically, the CNF leverages its high-efficiency compression capabilities to encode high-dimensional, intricate scientific data into a compact latent space, and simultaneously, an unconditional diffusion model operates in the CNF-encoded latent space, effectively generating new spatiotemporal sequences in a scalable manner. This unique integration catalyzes the formation of a novel class of latent diffusion models for space-time generation, marking a significant advancement in the field of generative modeling.

CoNFiLD has demonstrated its proficiency in generating a broad spectrum of turbulent flows across complex and irregular domains, successfully capturing intricate chaotic dynamics and turbulent phenomena. Moreover, CoNFiLD offers versatile zero-shot conditional generation capabilities, making it highly applicable to real-time data assimilation or scalable inverse problems in a variety of scientific and engineering applications, such as sensor-based flow reconstruction, data restoration, and super-resolution data enhancement, all without the necessity for model retraining. Additionally, our model exhibits robust performance and superior computational/memory efficiency compared to classic numerical simulations and other state-of-the-art generative AI techniques. While CoNFiLD is primarily showcased on large-scale turbulence generation in this study, its can be naturally applied to modeling general spatiotemporal dynamics across various scientific domains.

\section*{Acknowledgment}
The authors would like to acknowledge the funds from Office of Naval Research under award numbers N00014-23-1-2071 and National Science Foundation under award numbers OAC-2047127. XF would also like to acknowledge the fellowship provided by the Environmental Change Initiative and Center for Sustainable Energy at University of Notre Dame.

\section*{Compliance with Ethical Standards}
Conflict of Interest: The authors declare that they have no conflict of interest.

\bibliographystyle{elsarticle-num}
\bibliography{ref,myPub}

% \appendix
\end{document}